\documentclass[final,leqno]{siamltex}

\usepackage{graphicx}
\usepackage{latexsym}
\usepackage{amssymb}
\usepackage{amsmath}
\usepackage{amsmath,empheq}
\usepackage{booktabs}
\usepackage{subfig}
\usepackage[labelfont=bf,textfont=it,belowskip=0pt,aboveskip=5pt,tableposition=top]{caption}
\usepackage{enumerate}
\usepackage{multicol}
\usepackage{rotating}
\usepackage{pdflscape}
\DeclareMathSizes{12}{8}{7}{4}
\usepackage{nicefrac}
\usepackage{mathtools}
\usepackage{svg}
\usepackage{sidecap}
\usepackage{siunitx}
\usepackage{placeins}
\usepackage{comment}
\usepackage{sidecap}

\usepackage{xargs}                      

\usepackage[sc]{mathpazo}
\usepackage{algpseudocode}
\usepackage{algorithm}
\usepackage[sort,compress]{cite}
\usepackage{multirow}
\usepackage{makecell}
\usepackage{textcomp}
\usepackage{paralist}
\usepackage{cancel}
\usepackage{color, colortbl}
\usepackage{geometry}
\usepackage[colorlinks=true,
citecolor=blue,
filecolor=black,
linkcolor=blue,
urlcolor=blue]{hyperref}
\usepackage{longtable}
\usepackage{graphbox}
\usepackage{xspace}
\graphicspath{{figures/}{tables/}}

\newcommand{\MATLAB}{\textsc{Matlab}\xspace}

\newcounter{runidnum}

\newcommand{\highlightrow}{\rowcolor{gray!20}}

\newcommand{\figref}[1]{Fig.~\ref{#1}}
\newcommand{\tabref}[1]{Tab.~\ref{#1}}
\newcommand{\secref}[1]{\S\ref{#1}}
\renewcommand{\algref}[1]{Alg.~\ref{#1}}

\newcolumntype{R}{>{\columncolor{gray!20}}r}
\newcolumntype{L}{>{\columncolor{gray!20}}l}

\newcommand{\elltwoT}{\ensuremath{\mu_{T,L_2}}} 
\newcommand{\elltwoIC}{\ensuremath{\mu_{0,L_2}}} 
\newcommand{\elltwoICP}{\ensuremath{\mu_{\vect{p},\ell_1}}} 
\newcommand{\relG}{\ensuremath{\|\vect{g}\|_{\text{rel}}}} 

\newcommand{\mpPatientBrain}[1][]{\ifthenelse { \equal {#1} {} }
    { \vect{m}_D }   
    { \vect{m}_{D,#1} } }  

\newcommand{\mpWarpedAtlasBrain}[1][]{\ifthenelse { \equal {#1} {} }
    { \vect{m}_A^{(1,1)} }   
    { \vect{m}_{A,#1}^{(1,1)} }}   

\newcommand{\ns}[1]{\ensuremath{\mathbf{#1}}}

\newcommand{\idiv}{\ensuremath{\mbox{div}\,}}
\newcommand{\igrad}{\ensuremath{\nabla}}

\newcommand{\p} {\partial}

\newcommand{\vect}[1]{\boldsymbol{#1}} 
\newcommand{\mat}[1]{\boldsymbol{#1}}  

\newcommand{\bipa}{\begin{inparaenum}[(\itshape i\upshape)]}
\newcommand{\eipa}{\end{inparaenum}}

\newcommand{\bipasub}{\begin{inparaenum}[(\itshape a\upshape)]}
\newcommand{\eipasub}{\end{inparaenum}}

\definecolor{sred}{cmyk}{0.01,0.98,0,0.2} 
\reversemarginpar
\newcommand{\mmargin}[1]{{\marginpar{\em\tiny #1}}}\renewcommand{\mmargin}[1]{}

\title{Where did the tumor start? An inverse solver with sparse localization for tumor growth models}

\author{
	Shashank Subramanian\thanks{Oden Institute for Computational Engineering and Sciences, University of Texas at Austin,
		201 E. 24th Street, Austin, Texas, USA
		({\tt shashank@ices.utexas.edu})}
	\and
	Klaudius Scheufele\thanks{Institute for Parallel and Distributed Systems, Universit\"at
        Stuttgart, Universit\"atsstra{\ss}e 38, Stuttgart, Germany
        ({\tt klaudius.scheufele@ipvs.uni-stuttgart.de})}
        \and
        Miriam Mehl\thanks{Institute for Parallel and Distributed Systems, Universit\"at
        Stuttgart, Universit\"atsstra{\ss}e 38, Stuttgart, Germany
        ({\tt miriam.mehl@ipvs.uni-stuttgart.de})}
    \and
    	George Biros\thanks{Oden Institute for Computational Engineering and Sciences, University of Texas at Austin,
    		201 E. 24th Street, Austin, Texas, USA
    		({\tt biros@oden.utexas.edu})}
 }
\begin{document}

\maketitle

\begin{abstract}
We present a  numerical scheme for solving an inverse problem for parameter estimation in tumor growth models for glioblastomas, a form of aggressive primary brain tumor.  The growth model is a reaction-diffusion partial differential equation (PDE) for the tumor concentration. We use a  PDE-constrained optimization formulation for the inverse problem.  The unknown parameters are the reaction coefficient  (proliferation), the diffusion coefficient (infiltration), and the initial condition field for the tumor PDE. Segmentation of Magnetic Resonance Imaging (MRI) scans drive the inverse problem where segmented tumor regions serve as partial observations of the tumor concentration. Like most cases in clinical practice, we use data from a single time snapshot. Moreover, the precise time  relative to the initiation of the tumor is unknown, which poses an additional difficulty for inversion. We perform a frozen-coefficient spectral analysis and show that the inverse problem is severely ill-posed. We introduce a biophysically motivated regularization on the structure and magnitude of the tumor initial condition. In particular, we assume that the tumor starts at a few locations (enforced with a sparsity constraint on the initial condition of the tumor) and that the initial condition magnitude in the maximum norm is equal to one.  We solve the resulting optimization problem using an inexact quasi-Newton method combined with a compressive sampling algorithm for the sparsity constraint. Our implementation uses PETSc and AccFFT libraries. We conduct numerical experiments on synthetic and clinical images to highlight the improved performance of our solver over a previously existing solver that uses standard two-norm regularization for the calibration parameters. The existing solver is unable to localize the initial condition. Our new solver can localize the initial condition and recover infiltration and proliferation. In clinical datasets (for which the ground truth is unknown), our solver results in qualitatively different solutions compared to the two-norm regularized solver.

\end{abstract}


\begin{AMS}
35K40, 49M15, 49M20, 65K10, 65N35, 65Y05, 92C50
\end{AMS}
{\textit{\footnotesize{\hspace{-0.5em}(Preprint submitted to Journal of Inverse Problems)}}}

\pagestyle{myheadings}
\thispagestyle{plain}
\markboth{SUBRAMANIAN, SCHEUFELE, MEHL, BIROS}{An inverse solver with sparse localization for tumor growth models}

\section{Introduction}
\label{sec:intro}
  The integration of biophysical macroscopic brain tumor growth models with Magnetic Resonance Imaging (MRI) scans is an important tool in clinical cancer research. Phenomenological growth models capture phenomena like tumor cell invasion and proliferation and can be used to compute patient-specific biomarkers. Such biomarkers can potentially facilitate diagnosis (e.g., tumor grading), prognosis (e.g., predicting recurrence and survival), and treatment (e.g., pre-operative planning and radiotherapy). Additionally, biophysical models can advance our understanding of the disease by using imaging data to test model-driven hypotheses on disease progression and  treatment. For example, one such hypothesis is that the location of tumor initiation relates to distinct tumor mutations, thereby paving the way to molecularly-informed models of tumor growth~\cite{akbari-e18}. The main problem we focus here is whether we can localize the initiation of the tumor while simultaneously estimating physics-based parameters that characterize infiltration (spreading of the tumor) and proliferation (growth).

We use a single-species, reaction-diffusion model of tumor growth (described in detail in~\secref{sec:forward}),
\[
\partial_t c(\vect{x},t) = \kappa {\mathcal D} c(\vect{x},t) + \rho {\mathcal R} c(\vect{x},t), \quad  t\in[0,T], \quad \mbox{with~initial~condition~} c(\vect{x},0),
\]
 where $c(\vect{x},t)$ is the tumor volume fraction (or concentration, assuming constant density), $\kappa$ is a scalar diffusion parameter, $\rho$ is a scalar reaction (or growth) parameter,  ${\mathcal D}$ is a diffusion operator (with free-flux boundary conditions), and ${\mathcal R}$ is an non-linear logistic growth operator.  This is one of the most common tumor growth models owing to its simplicity (it has just two parameters and the initial condition) and good performance in capturing the basic characteristics of brain tumor growth observed in MRI scans~\cite{swanson2000quantitative,rockne-e19}. The parameters of this model are the reaction coefficient $\rho$ (modeling proliferation), the diffusion coefficient $\kappa$ (modeling invasion), and the initial condition $c(\vect{x},0)$. For the inverse problem (described in~\secref{sec:inverse}), we assume that we are given \emph{partial} observations of $c(\vect{x},T)$ at some, unknown relative to the tumor initiation,  time $T$ and we seek to reconstruct  $\kappa$, $\rho$, and $c(\vect{x},0)$, which are unknown. The time horizon $T$ is also unknown but a simple calculation (see~\secref{sec:formulation}) shows that among $\kappa$, $\rho$, and $T$, only two are independent.  Hence, we set  $T=1$ and invert for the remaining parameters.  Notice that $c(\vect{x},0)$ is a field. In our formulation, we parameterize it as a finite sum of radial functions. This re-parameterization is not only for regularization but mainly to allow for $\ell_1$ sparsity constraints and can be viewed as a modeling assumption.

 The data used for the inverse problem are partial observations of the tumor concentration \emph{at a single time point}. This scenario corresponds to the  clinical setting  in which we seek  to estimate the tumor grade and aggressiveness from a single-time-point scan (as opposed to the case in which we have a series of scans taken at different points in time). Indeed, when a patient comes with symptoms, the tumor is typically quite large and treatment (surgery, chemotherapy, or radiation) starts immediately. Hence, it is not practical to assume multiple time snapshots of the tumor growth because otherwise we would have to model treatment (which comes with its own modeling challenges and introduces additional unknown parameters). Given this setup, the inverse problem is formulated as a PDE-constrained optimization problem. We seek to estimate the model parameters that minimize the mismatch between the model-generated tumor and the partially observed tumor concentration. This problem is highly ill-posed. Notice that just inverting for the initial condition problem is exponentially ill-posed (see~\secref{sec:1d}).  The inherent non-linearity in the tumor model and the strong ill-posedness of the associated inverse problem make this effort a computationally challenging task. In this paper, we discuss this ill-posedeness, and we propose a set of algorithms for its numerical solution.

\subsection{Contributions}
The main innovation in the paper is the formulation of an $\ell_1$-constrained inverse problem for the initial condition, and two model parameters, the reaction $\rho$ and the diffusion $\kappa$. Our contributions are summarized below.
\begin{itemize}
\item We formulate the inverse problem as a PDE-constrained optimization problem~(see \secref{sec:formulation}). We perform an analysis of the linearized operator to demonstrate the ill-posedness and the need for a specific regularization~(see \secref{sec:1d}). Although an $\ell_2$ regularization (for all parameters) removes the ill-posedness,  it does not result in good localization of the initial condition and tumor parameters. For this reason, we introduce an $\ell_1$ sparsity constraint to enable sparse\footnote{We denote an initial condition parameterization with only few non-zero coefficients as sparse.} reconstruction for the initial condition, combined with a maximum norm condition for $c(\vect{x},0)$. This allows us to reconstruct the model parameters $\kappa$ and $\rho$.  This regularization is biophysically motivated.
\item We introduce a novel numerical scheme to simultaneously invert for all model parameters while preserving the sparse content of the initial tumor condition. We employ a quasi-Newton optimization solver along with a reduced-space parameterization of the tumor initial condition~(see~\secref{sec:numerics}). The scheme alternates between an $\ell_2$ regularized solve and an $\ell_1$ projection. The underlying linear and non-linear solves are based on preconditioned inexact Krylov solves accelerated with PETSc and AccFFT libraries that support distributed-memory parallelism and enable fast solves for the challenging derivative-based 3D problem.
\item We test our scheme on 3D clinical and  synthetic datasets. The synthetic datasets resemble the complexity in tumor patterns observed in MRI scans and are used to verify our algorithm since we know the ground truth. With the clinical images,  we examine the prediction error and compare with a classical $\ell_2$ regularization scheme.  We report results that highlight the performance improvement of our solver over existing inversion strategies in estimating $c(\vect{x},0)$, $\kappa$, and $\rho$. We further demonstrate the robustness of our solver by testing it on noisy data and partial observations. This inversion setup is related to the clinical setting because the extent of tumor invasion is not fully observable in MRI scans.
\end{itemize}

\subsection{Limitations}
 While the reaction-diffusion model can predict the whole tumor region, it does not distinguish between the different tumor phenotypes observed in imaging data (such as enhancing and necrotic tumor regions). Further, it does not account for any mechanical deformations on the ventricles and surrounding healthy tissue due to tumor growth. We have developed more complex models that do so~\cite{subramanian-gholami-biros19}, but their integration with the inverse problem is beyond the scope of this paper.
We do not have direct observations of the tumor concentration, we only have the segmentation of the MRI scan into healthy tissue, edema, enhancing tumor, non-enhancing tumor, and necrotic tumor. In this paper, we ignore these glioma-specific details, and  assign a volume fraction one to regions of enhancing, non-enhancing, or necrotic tumor.
  Our formulation requires the healthy patient brain scan, which is typically not available. The  standard approach in medical imaging is to resort to methods such as image registration to construct a healthy atlas that resembles the observed healthy structures of  the patient scans. This can be inaccurate for very large tumors.  We are currently working on incorporating the ideas presented here with our previous work~\cite{Scheufele:2017a}, in which we coupled diffeomorphic image registration with tumor inversion for this purpose.

\subsection{Related work}
Although there has been a lot of work on forward problems for tumor growth, there has been less work on inverse problems. The latter has different aspects. The first is the underlying biophysical model. The second is the inverse problem setup, observation operators and the existence of scans at multiple points, the noise models, inversion parameters and constraints. And the third one is the solution algorithm.

Regarding the underlying model, like us, most researchers focus on parameter calibration of a handful of model parameters using single-species reaction-diffusion equations~\cite{clatz2005realistic,hogea2007modeling, hogea2008brain,jbabdi2005simulation,konukoglu2010image, mang2012biophysical, swanson2000quantitative,swanson2002virtual, rekik2013tumor}. While more complex models describing processes like mass effect, angiogenesis and chemotaxis~\cite{saut2014multilayer, pham2012density, Swanson:2011a, Hogea:2007b, hormuth2018} exist, they have not been considered for calibration due to theoretical and computational challenges. However, several groups, including ours, are working to address these challenges.

Regarding  the inverse problem setup, in most studies the authors use scans from two or more time points and assume that the tumor concentration is fully observed. In that case the inverse problem is much more tractable since we need to invert for just $\kappa$ and $\rho$.  Some authors use manual calibration, which is possible for a handful of parameters~\cite{clatz2005realistic, swanson2002virtual}. But most research groups resort to derivative-free optimization algorithms including Bayesian methods~\cite{Chen:2012b,Konukoglu:2010a,Mang:2014a,Mang:2012a,Mi:2014a,Wong:2017a, HawkinsDaarud:2013b,Le:2015a,Le:2017a}. In~\cite{konukoglu2010image}, the authors use the traveling wave approximation to the reaction-diffusion model to estimate the diffusivity by fixing the reaction coefficient and the source seed. They explain that the non-uniqueness of the solution renders the inversion of all parameters infeasible for their  model. Derivative-based optimization methods have been employed in~\cite{Colin:2014a,Hogea:2008b,Gholami:2016a,Knopoff:2013a,Mang:2014a,mang2012biophysical,Quiroga:2015a, Scheufele:2017a}.

Now we discuss, in some detail, attempts to reconstruct the initial condition of the tumor using a single scan.  In our  previous works~\cite{Gholami:2016a, Scheufele:2017a}, we employed an $L_2$ regularization scheme for $c(\vect{x},0)$ and we inverted for $\kappa$ assuming known $\rho$. We used reduced-space parallel algorithms allowing for the calibration of highly complex tumor shapes in reasonable time. There have been a few studies exploring the identifiability of all parameters and their correlation with the tumor source. The traveling wave approximation (using an Eikonal equation) from~\cite{konukoglu2010image}, is extended in~\cite{REKIK2013238} to estimate a localized initial condition and the diffusion coefficient. To our knowledge, this is the only attempt to localize the initial condition of the tumor. Its advantage is that it is simple to implement and quite robust since it circumvents non-convexity. But it is limited to the specific Eikonal equation for the propagation of the tumor front, which is then time-reversed to localize the tumor. Unlike our scheme, this inversion scheme cannot be used with more realistic biophysical models. In~\cite{Jaroudi2019}, the authors use the 3D non-linear model and attempt to identify  $c(\vect{x},0)$ by fixing the growth rate and the diffusion coefficient to set values. The scheme is essentially an unregularized gradient descent and assumes noise-free, full observations of $c(\vect{x},1)$. The authors consider only synthetic experiments and assume $\kappa$ and $\rho$ are the ground truth. In this idealistic setup everything works well and one can localize $c(\vect{x},0)$. As we will see, in the presence of noise, partial observations, and uncertainty in $\kappa$ and $\rho$,  such a formulation fails.

Let us also mention that, in general (unrelated to tumor modeling), there is little work on sparsity constraints for  initial condition inverse problems for reaction-diffusion PDEs. While some groups have used sparsity promoting approaches to estimate sources in the heat equation in a 1-D and 2-D setting~\cite{Li:2014, Flinth:2019}, there is not much work on sparse localization for the non-linear reaction-diffusion equations.  In conclusion, there is no prior work on a generalizable scheme for inverting for a sparse initial condition for the tumor in addition to model parameters.

Regarding numerical methods for $\ell_1$ regularized inverse problems, there is a vast amount of work in literature. Most algorithms are based on compressive sampling pursuit~\cite{Tropp:2007, Donoho:2012, Needell:2010, NEEDELL2009301, Blumensath:2009}, $\ell_1$ convex optimization approaches using methods like interior points~\cite{Candes:2006, Kim:2007}, projected gradient algorithms~\cite{Fig:2007}, or iterative shrinkage algorithms~\cite{Khor:2019 ,Beck:2009, Gong:2013}.  We base our algorithm  on the enhanced greedy pursuit strategy described in~\cite{NEEDELL2009301},  which yields optimal theoretical guarantees on convergence rates for quadratic mismatch functions. However, our overall problem is ill-posed and non-convex. To our knowledge, fast algorithms with theoretical guarantees for problems like ours do not exist.

\subsection{Outline}
In \secref{sec:methods}, we present the forward problem and the  mathematical formulation of the parameter estimation inverse problem. We explore the invertibility of the unknown parameters of the growth model in a simplified 1D setting and introduce new assumptions to solve the parameter estimation problem. In \secref{sec:numerics}, we provide details regarding the algorithm and numerical schemes used. Finally, we conduct numerical experiments to assess the performance of our solver in \secref{sec:results}.


\section{Methods}\label{sec:methods}
We introduce some notation and subsequently present the forward problem and the mathematical formulation of the inverse problem below.
\subsection{ Formulation}
\label{sec:formulation}
\begin{figure}
	\centering
	\subfloat[]{\includegraphics[width=0.216\textwidth]{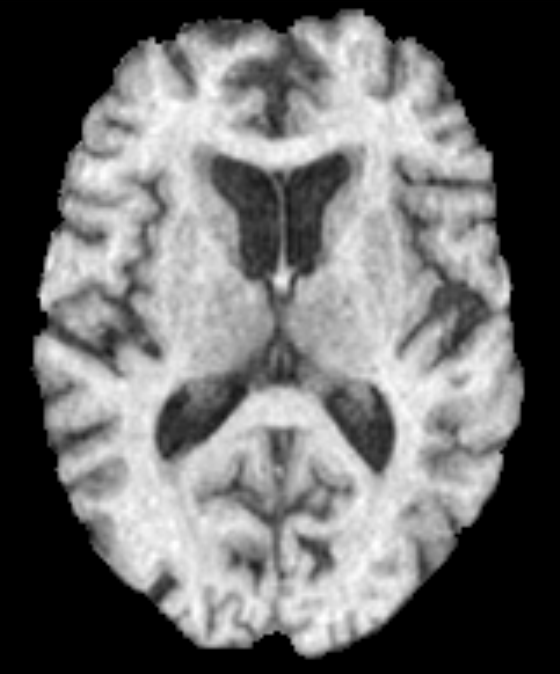}}\hspace{0.1em}
	\subfloat[]{\includegraphics[width=0.216\textwidth]{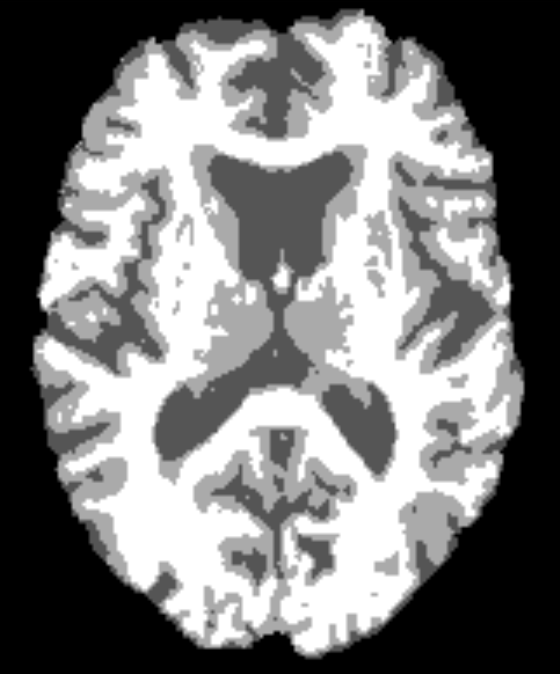}}\hspace{0.1em}
	\subfloat[]{\includegraphics[width=0.2099\textwidth]{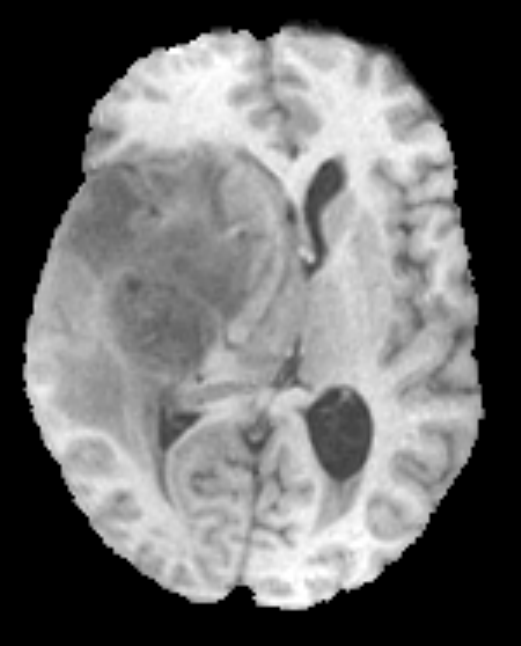}}\hspace{0.1em}
	\subfloat[]{\includegraphics[width=0.2099\textwidth]{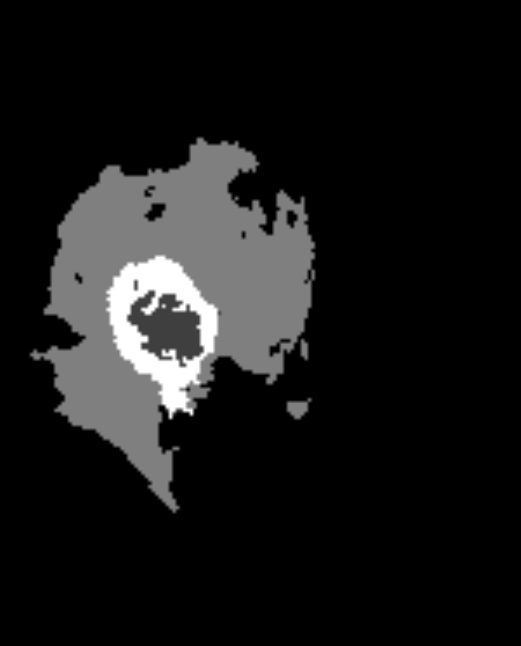}}\hspace{0.1em}
	\caption{\textit{Imaging data which show an axial slice of (a) healthy T1 MRI scan, (b) segmentation of the healthy brain into white matter (white), gray matter (light gray), and cerebrospinal fluid-filled ventricles (dark gray), (c) patient T1 MRI scan diagnosed with glioblastoma, and (d) segmentation of the patient tumor sub-structures into enhancing tumor (white), necrotic tumor (dark gray), and edema (light gray). The patient scan and segmentation are taken from the 2018 Multimodal Brain Tumor Segmentation challenge training dataset~\cite{Bakas17}, while the healthy brain scan and segmentation are obtained from the GLISTR dataset~\cite{gooya-biros-davatzikos-e12}. In our formulation, we assume the tumor core (enhancing and necrotic tumor) to be the tumor segmentation.}\label{fig:segmentations}}
\end{figure}

We have briefly discussed the forward model for the tumor concentration $c(\vect{x},t)$. It involves two operators, the diffusion operator ${\mathcal D}$ and the reaction operator ${\mathcal R}$. In order to define these operators, we need to take into account the material heterogeneity of the brain tissue. For example, tumors do not grow inside the ventricles, so the reaction and diffusion coefficients should be set to zero. Also, we need to account for the fact that we do not have the precise tissue type, but only probability maps related to the MRI scans. These probability maps are obtained by statistical analysis of manual or machine-learning based segmentations~\cite{gooya2011deformable}.  So, in our setting, a brain is described by separate probability maps for each brain tissue type, namely gray matter, white matter, and cerebrospinal fluid (CSF; which includes ventricles). We represent such probability maps as a vector field:
\begin{equation}
\vect{m}(\vect{x}) =
\left(
\begin{array}{c} m_g(\vect{x}) \\ m_w(\vect{x}) \\ m_f(\vect{x}) \end{array}
\right)
\in\ns{R}^3,
\label{e:vec_m}
\end{equation}
where $\vect{x} = (x,y,z)$ are the spatial coordinates and $m_g(\vect{x})$, $m_w(\vect{x})$ and $m_f(\vect{x})$ denote the gray matter, white matter and CSF probability maps, respectively. These probability maps are obtained from binary segmentation labels for each tissue type of a healthy brain MRI scan.\footnote{In a pre-processing step, the binary labels are smoothed and re-scaled to probabilities.} Our data $d(\vect{x})$ is derived from actual patient MRI tumor segmentations.\footnote{We assign $d(\vect{x}) = 1$ to the voxels that have been identified as tumor. More complex scenarios are possible, for example see~\cite{Castillo60}, which uses apparent diffusion coefficient (ADC) values to identify tumor tissue.} Figure~\ref{fig:segmentations} shows an example of a healthy brain MRI scan (and its associated segmentation) and a patient brain MRI scan (and its associated tumor segmentation). As we discussed the healthy patient brain scan is rarely available in practice. There have been many approaches to address this, such as using a statistical average of many individual brains or image registration methods to approximate the healthy patient brain (see~\cite{Scheufele:2017a} for details). For the most part of this paper, we assume that the healthy patient brain segmentation is given. For our test on an actual clinical scan, we use an atlas-based segmentation in~\secref{sec:rt}.

\subsubsection{Forward problem}\label{sec:forward}
The forward problem is a single-species, reaction-diffusion partial differential Eq.~\cite{hogea2007modeling, hogea2008brain, jbabdi2005simulation, konukoglu2010image, mang2012biophysical, swanson2000quantitative, murray1989mathematical}.  In this model  we group all the MRI-observed tumor phenotypes into a single one. (For gliomas, these phenotypes are typically enhancing, non-enhancing, and necrotic. We do not include edema in the tumor although it is quite typical to have significant tumor infiltration in the edema region.)  As mentioned in the introduction, the time horizon (the time since tumor initiation at which we observe the tumor concentration) is unknown. However, we can non-dimensionalize the PDE and use a normalized time horizon equal to one for tumor growth so that $t \in [0, 1]$. We discuss this further below. Let $\tilde{t} = t/T$ be the normalized time. Then, the PDE can be written as:
\begin{equation}
    \p_{\tilde{t}} c - T \kappa  \mathcal{D} c  -  T \rho  \mathcal{R} c =  0. 
\end{equation}
Denoting $\tilde{\kappa} = \kappa T$ and $\tilde{\rho} = \rho T$, we observe that the normalization of time is equivalent to a re-scaling of the reaction and diffusion coefficients. In the remaining of the paper, we use $t, \kappa, \rho$  instead of $\tilde{t}, \tilde{\kappa}, \tilde{\rho}$ to represent the re-scaled parameters. This normalization is because we cannot possibly invert for the three parameters $\rho$, $\kappa$, and $T$, since only two are independent. So, we set $T=1$ and we consider the reconstruction of $\rho$ and $\kappa$ (in addition to the initial condition).
The brain scan is re-scaled to be in $\Omega = [0, 2\pi]^3$. The non-dimensionalized forward problem is given as:
\begin{align}
& &\p_t c - \kappa \mathcal{D} c  - \rho \mathcal{R} c&=  0 & &\mbox{in}~U = \Omega \times (0,1],
\label{eq:fwd} \\
& & c(0)  &= g& &\mbox{in}~\Omega,
\label{eq:fwd_initial}
\end{align}
where $c = c(\vect{x},t)$ is the tumor volume fraction  (or mass concentration assuming constant density), $c(0) = c(\vect{x}, 0)$ is the initial tumor concentration set to $g = g(\vect{x})$, $\mathcal{D}$ is the diffusion operator, $\kappa$ is a scalar diffusion coefficient, $\rho$ is a scalar growth coefficient, and $\mathcal{R}$ is the reaction operator. We use periodic boundary conditions on $\partial\Omega$. In reality, we have  no-flux boundary conditions on the boundary of the skull and CSF boundaries. These no-flux conditions are approximated using a penalty method, the details are discussed in~\cite{Gholami:2016a}. The approximation amounts to having very small diffusivity in the ventricles and the image space surrounding the brain.  We define $\mathcal{D}$ to be an inhomogeneous isotropic diffusion operator given by
\begin{equation}\label{eq:diff_op}
\mathcal{D}c = \idiv k_{\vect{m}} \igrad c,
\end{equation}
where $k_{\vect{m}}(\vect{x})$ is defined as:
\begin{equation}
k_{\vect{m}}(\vect{x}) =  \kappa_{g} m_{g}(\vect{x}) + \kappa_{w} m_{w}(\vect{x})  \label{eq:diff_coeff},
\end{equation}
where $\kappa_g$ and $\kappa_w$ are scalar parameters which enforce inhomogeneity of the diffusion rate in the gray matter and white matter, respectively. Similarly, we define the reaction operator $\mathcal{R}$ as:
\begin{equation}\label{eq:reac_op}
\mathcal{R}c =  \rho_{\vect{m}} c(1 - c),
\end{equation}
with $\rho_{\vect{m}}(\vect{x})$ defined as:
\begin{equation}
\rho_{\vect{m}}(\vect{x}) = \rho_{g} m_{g}(\vect{x}) + \rho_{w} m_{w}(\vect{x})   \label{eq:reac_coeff},
\end{equation} 
where $\rho_g$ and $\rho_w$ are scalar parameters which control the spatial variability of the growth rate in the gray matter and white matter, respectively. The tumor does not grow or diffuse in the cerebrospinal fluid-filled ventricles (CSF).

\subsubsection{Inverse problem}\label{sec:inverse}
Our formulation of the inverse problem requires the definition of the observation operator,  the mismatch function, and the initial condition parameterization.  Existing MRI technologies (the predominant imaging modality used in neuro-oncology) do not provide the ability to directly observe the extent of tumor concentration everywhere in the parenchyma. For this reason, we introduce an observation operator $\mat{O}(\vect{x})$ that is determined by the clearly observable tumor margin. We also assume that in that region the tumor concentration is equal to one. We now present the formulation for the inverse problem and explain its components below.

The inverse tumor growth problem is given by the following minimization problem:
\begin{subequations}  \label{eq:tumor_opt}
\begin{gather}\label{eq:tumor_objective}
\min_{\vect{p},\kappa, \rho}~ \mathcal{J} (\vect{p},\kappa, \rho) :=
\frac{1}{2} \| \mat{O}c(1) - d \|_{L_2(\Omega)}^2 + \frac{\beta}{2} \| \boldsymbol{\phi}^T \vect{p} \|_{L_2(\Omega)}^2
\end{gather}

\noindent subject to the reaction-diffusion forward problem $\mathcal{F}(\vect{p}, \kappa, \rho)$:
\begin{align}
& &\p_t c - \kappa \mathcal{D} c  - \rho \mathcal{R} c&=  0 & &\mbox{in}~U = \Omega \times (0,1],
\label{eq:tumor:state} \\
& & c(0)  &= \boldsymbol{\phi}^T \vect{p} & &\mbox{in}~\Omega.
\label{eq:tumor:initcond}
\end{align}

\end{subequations}
Here, we have introduced a parameterization for the tumor initial condition $c(0) = c(\vect{x}, 0)$ as $\boldsymbol{\phi}^T(\vect{x})\vect{p}$, where
$\vect{p}$ is an $m$-dimensional vector, $\phi_i(\vect{x}) = \phi_i(\vect{x}-\vect{x_i})$ is a radial function centered at point $\vect{x}_i$ (one of the grid points after spatial discretization), and $\boldsymbol{\phi}(\vect{x}) = \{\phi_i(\vect{x}) \}_{i = 1}^m$, i.e.,
\begin{equation}\label{e:phi}
c(\vect{x}, 0) = \boldsymbol{\phi}^T(\vect{x}) \vect{p} = \sum_{i=1}^{m} \phi_i(\vect{x}) p_i, \quad  \vect{p} \in\ns{R}^{m}.
\end{equation}
This parameterization is mainly a modeling assumption. We assume that the spatial distribution of the tumor at its very early stages is given by $\phi_i$ for some $i$. There is no clear understanding on how tumors start, but the general consensus is that it starts at a single cell due to some mutation. At this resolution the continuum hypothesis breaks down and it is unclear how to couple such a single cell mutation to the PDE model. Instead, we use the PDE model after the tumor has grown sufficiently large while remaining localized. We label this time as zero and assume that the tumor volume fraction can be approximated well by  $\phi(\vect{x}-\vect{x}_i)$, where $\vect{x}_i$ is the center of mass of this tumor. However, we allow a small number of $\vect{x}_i$  points to be activated. This is to admit more complex shapes for the initial condition and also to compensate for the modeling errors in our tumor growth PDE. 

The objective function minimizes the mismatch between the simulated tumor at $t = 1$ and data on all observation points using the standard $L_2$ distance measure on square integrable functions. We balance this mismatch with an $L_2$ regularization on the inverted initial condition with regularization parameter $\beta$.

\begin{table}
	\centering
	\caption[Summary of Notation]{\label{tab:notation}Notation for the inverse tumor problem formulation.}
	\setlength\tabcolsep{5pt}\small
	\begin{tabular}{lc}
		\toprule
		Description              & Notation            \\
		\midrule
		\textit{spatial coordinates}			& $\vect{x} = (x,y,z)$ \\
\highlightrow		\textit{time}									& $t$ \\
		\textit{healthy tissue probability maps} (see Eq.~\eqref{e:vec_m})     & $\vect{m}(\vect{x})$    \\
\highlightrow		\textit{patient tumor from imaging} (see Eq.~\eqref{eq:tumor_objective})     & $d(\vect{x})$        \\
		\textit{initial tumor concentration} (see Eq.~\eqref{eq:tumor:initcond}) & $c(\vect{x},0)$  \\
\highlightrow		\textit{simulated tumor concentration at time $t$}  (see Eq.~\eqref{eq:tumor:state})   & $c(\vect{x},t)$           \\
		\midrule
		\textit{reaction operator}	(see Eq.~\eqref{eq:reac_op})	&	$\mathcal{R}$ \\
\highlightrow		\textit{scalar reaction coefficient}	(see Eq.~\eqref{eq:reac_coeff})	&	$\rho$ \\
		\textit{diffusion operator}	(see Eq.~\eqref{eq:diff_op})	&	$\mathcal{D}$ \\
\highlightrow		\textit{scalar diffusion coefficient} (see Eq.~\eqref{eq:diff_coeff})	&	$\kappa$ \\
		\textit{parameterization basis functions}	(see Eq.~\eqref{e:phi})										&		$\boldsymbol{\phi}(\vect{x}) = \{\phi_i(\vect{x})\}_{i = 1}^m$ \\
\highlightrow		\textit{parameterization coefficients in $\ns{R}^{m}$}	(see Eq.~\eqref{e:phi})		& 		$\vect{p}$ \\
		\textit{observation operator}	(see Eq.~\eqref{e:obs})		& 		$\mat{O}(\vect{x})$ \\
\highlightrow		\textit{tumor concentration detection threshold} (see Eq.~\eqref{e:obs})			& 	$c_d$ \\
		\textit{regularization parameter} (see Eq.~\eqref{eq:tumor_objective})     & $\beta$        \\
		\bottomrule
	\end{tabular}
\end{table}

 Similar to the works of~\cite{Gholami:2016a}, we observe all tumorous data points with concentration above a detection threshold ($c_d$). We define our observation operator $\mat{O}$ as follows:
\begin{equation}\label{e:obs}
	\mat{O}(\vect{x}) =
	\begin{cases}
	1,& \text{if } d(\vect{x}) \geq c_d\\
	0,              & \text{otherwise.}
	\end{cases}
\end{equation}
In our synthetic examples, we know the ground truth $c$. In clinical images, we use the segmentation probability maps as a proxy for $c$.  \tabref{tab:notation} summarizes the notations. Next, we discuss the difficulty of inverting for   $\vect{p}$, $\kappa$ and  $\rho$ simultaneously. To do that, we  use a simplified analytical model in one dimension.

\subsection{Parameter inversion analysis for a simple linear analytical model}
\label{sec:1d}
There have been several studies that outline the infeasibility of inverting for all parameters of a reaction-diffusion system. In particular, the authors in~\cite{konukoglu2010image} note the non-uniqueness of the solution of the forward model with respect to different parameter configurations thereby rendering the inverse problem ill-posed. Here, we bolster this intuitive explanation through an analytical investigation of a simplified model which underpins our strategy in circumventing this issue.

We consider a 1D \emph{linear} reaction-diffusion model with constant coefficients on the domain $\omega = [0,\pi]$ with periodic boundary conditions as our growth model. The forward model is given by:
\begin{equation}
	\begin{aligned} \label{eq:1D_linear_model}
		\partial_t c - \rho c - \kappa\partial^2_x c &= 0 & &\mbox{in}~ \omega \times (0,1], \\
		\left. \partial_x c \right|_{x = 0, \pi}&= 0, \\
		c(x,0) - g(x) &= 0 & &\mbox{in}~\omega.
	\end{aligned}
\end{equation}
where $c(x, t) = c$ is the tumor volume concentration, $\rho$ is the reaction coefficient, and $\kappa$ is the diffusion coefficient.
Our goal is to invert for $g(x)$, $\rho$ and $\kappa$. Notice that, for a moment here, we omit the parameterization we introduced in the previous section for $c(x,0)$ and we just try to invert for a field $g(x)$. We can solve Eq.~\eqref{eq:1D_linear_model} analytically using separation of variables and we obtain
\begin{equation}
	c(x,t) = \sum_{n=0}^{\infty} g_n e^{(\rho - n^2\kappa)t}\cos(nx),
\end{equation}
where $g_n$ are the spectral cosine coefficients of the initial condition $g(x)$. We truncate this infinite sum to $N$ terms, where we assume that the solution is fully resolved for sufficiently large $N$. For our inverse problem, we assume band-limited data given by: $d(x) = \sum_{n=0}^{N} d_n \cos(nx)$. Now, we can formulate the parameter inversion as the following optimization problem for $\{g_n\}_{n=0}^N$, $\rho$ and $\kappa$ as:
\begin{equation}\label{eq:1D_opt}
	\min_{g_n, \rho, \kappa} \mathcal{J} := \frac{1}{2} \sum_{n = 0}^{N} \left(g_n e^{(\rho - n^2\kappa)} - d_n\right)^2.
\end{equation}
The first order optimality conditions to Eq.~\eqref{eq:1D_opt} are:
\begin{subequations}
\begin{align}
& &\partial_{g_n} \mathcal{J} &= 0:  & (g_n e^{(\rho - n^2\kappa)} - d_n)e^{(\rho - n^2\kappa)} &= 0, & \\
& &\partial_{\rho} \mathcal{J} &= 0: & \sum_{n = 0}^{N} (g_n e^{(\rho - n^2\kappa)} - d_n)g_ne^{(\rho - n^2\kappa)} &= 0 , & \\
& &\partial_{\kappa} \mathcal{J} &= 0: & -\sum_{n = 0}^{N} (g_n e^{(\rho - n^2\kappa)} - d_n)g_nn^2e^{(\rho - n^2\kappa)} &= 0. &
\end{align}
\end{subequations}
We observe that all equations are trivially satisfied when:
\begin{equation}
g_n e^{(\rho - n^2\kappa)} - d_n = 0. \label{eq:trivial_grad_condition}
\end{equation}
Eq.~\eqref{eq:trivial_grad_condition} is informative in two ways. First, we see that even if $\rho$ and $\kappa$ are known, inverting for $g_n$ is exponentially ill-posed. Indeed, $g_n = e^{-\rho} e^{n^2\kappa} d_n$. Any perturbation in $d_n$ is amplified exponentially in the frequency and thus, in practice, $g_n$ cannot be identified for large $n$.  But in our case, even small $n$ is problematic when $\rho$ and $\kappa$ are unknown. We cannot invert for $g_n$, $\rho$ and $k$ simultaneously since changing $\rho$ or $\kappa$ amounts to a scalar scaling of the initial condition. 

This ill-posedness persists even if we re-parameterize the initial condition as before using $m$ degrees of freedom:
\begin{equation}
	g(x) = \sum_{i = 0}^{m} \phi_i (x) p_i = \boldsymbol{\phi}^T(x) \vect{p},
\end{equation}
where $\phi_i$ are some pre-selected basis functions and $\vect{p} \in\ns{R}^{m}$ is the (low dimensional) coefficient vector. Taking the Fourier decomposition of $\phi_i$, we have: $\phi_i(x) = \sum_{n = 0}^{N}\phi_{in} \cos(nx)$, with spectral cosine coefficients $\phi_{in}$. Now we can write $g(x) = \sum_{n = 0}^{N} (\boldsymbol{\phi}_n^T \vect{p}) \cos(nx)$, with $\boldsymbol{\phi}_n = \{\phi_{in}\}_{i = 1}^m$.
Hence, the solution to the forward model is given by:
\begin{equation}
	c(x,t) = \sum_{n=0}^{N} \boldsymbol{\phi}_n^T \vect{p}  e^{(\rho - n^2\kappa)}\cos(nx),
\end{equation}
with the optimization problem as:
\begin{equation} \label{eq:optim_param}
	\min_{\vect{p}, \rho, \kappa} \mathcal{J} := \frac{1}{2} \sum_{n = 0}^{N} \left(\boldsymbol{\phi}_n^T \vect{p}  e^{(\rho - n^2\kappa)} - d_n\right)^2.
\end{equation}
Similar to the non-parametric case, it is not possible to invert for the initial condition and $\rho$ simultaneously as any perturbation in $\rho$ simply scales the initial condition. In contrast to $\rho$, a perturbation in $\kappa$ cannot be absorbed into the initial condition anymore due to the frequency dependency of the factor $e^{-n^2\kappa}$.

In order to recover the reaction coefficient, we make the following modeling assumption: $\max_x(g(x)) = 1$. Notice that, since $c(x,t)$ and $g(x)$ are volume fractions, their values are between zero and one. But here we assert something stronger. We assume that $t=0$ corresponds to the first time the tumor volume concentration reaches 100\% in some locations. This fixes the scale of the tumor initial condition and allows us to determine a reaction coefficient.  Our second modeling assumption is to use a Gaussian function for $\phi_i(x)$ (see~\secref{sec:numerics} for more details).  Also, notice that the equation for the diffusion is better defined. 

Hence, in summary, we can invert for $\kappa$ and the initial condition parameterization simultaneously, and can recover $\rho$ with an additional modeling assumption.

\subsection{Optimality conditions for the full 3D problem}
\label{sec:optm}
For the optimization, we additionally impose the bound constraints $\rho \geq 0$, $\kappa \geq 0$ and $\vect{p} \geq 0$ by definition of the forward problem. Hence, our formulation in Eq.~\eqref{eq:tumor_opt} is now constrained by:
\begin{align}
& & \vect{p} &\geq 0	& &\mbox{in}~\ns{R}^{m} \\
& & \kappa &\geq 0 & &\mbox{in}~\ns{R}  \label{eq:constraint_k}\\
& & \rho &\geq 0 & &\mbox{in}~\ns{R} \label{eq:constaint_rho}\\
& & \max (\boldsymbol{\phi}^T \vect{p}) &= 1  & &\mbox{in}~\Omega \label{eq:constaint_p}.
\end{align}

Eq.~\eqref{eq:constaint_p} is difficult to impose formally. So we will omit it from the optimality conditions and we will use it when we discuss the numerical scheme.  We introduce the Lagrangian and take variations with respect to $(c(\vect{x}, t), \alpha(\vect{x}, t), \vect{p})$, where $\alpha(\vect{x}, t)$ is the adjoint variable of $c(\vect{x},t)$.  This results in the following strong form of the \textbf{first order optimality conditions}.
%
\begin{subequations}\label{eq:first_order_opt}
\begin{align}
  \mbox{\textit{forward}} \hspace{105pt}\quad \quad
   \p_t c - \kappa \mathcal{D} c - \rho \mathcal{R}c &=  0  &&\mbox{in}~U, \label{eq:opt-cond:state} \\
 c(0) - \boldsymbol{\phi}^T \vect{p} &= 0 &&\mbox{in}~\Omega. \label{eq:opt-cond:state-init} \\
\mbox{\textit{adjoint} }  \hspace{67pt} \quad \quad 
 -\partial_t \alpha -  \kappa \mathcal{D} \alpha  - \rho \partial_{\vect{c}} (\mathcal{R}c) \alpha &= 0 & &\mbox{in}~\bar{U},
             \label{eq:opt-cond:adj} \\
                          \mat{O}^T(\mat{O}c(1) - d) + \alpha(1) &= 0 & &\mbox{in}~\Omega.
             \label{eq:opt-cond:tumor:adj-final} \\
 \mbox{\textit{gradient} } \hspace{10pt}\quad \quad
             \mathbf{g}_{\vect{p}} = \beta \left(\int_{\Omega} \boldsymbol{\phi} \boldsymbol{\phi}^T d\vect{x}\right) \vect{p} - \int_{\Omega} \boldsymbol{\phi} \alpha(0) d\vect{x} &= \vect{0} & &\mbox{in}~\Omega,
               \label{eq:opt-cond:inv} \\
{g}_{\kappa} = \int_0^1 \int_{\Omega} k_{\vect{m}} \left( (\igrad c)^T \igrad\alpha \right) \, d\vect{x}\,dt & = {0} &&\mbox{in}~\Omega,
             \label{eq:opt-cond:inv_diff}  \\
{g}_{\rho} = \int_0^1 \int_{\Omega} \rho_{\vect{m}} \left( -\alpha c (1 - c) \right) \, d\vect{x}\,dt & = {0} &&\mbox{in}~\Omega.
\label{eq:opt-cond:rho_diff}
\end{align}
\end{subequations}
The basic solution strategy is to iterate in $(\vect{p}, \kappa, \rho)$ variables using a bound constrained optimization algorithm in which at each iteration Eq.~(\ref{eq:opt-cond:state}, \ref{eq:opt-cond:state-init}) (\textit{forward}) and Eq.~(\ref{eq:opt-cond:adj}, \ref{eq:opt-cond:tumor:adj-final}) (\textit{adjoint}) are eliminated. Next, we describe the details of the numerical schemes for the forward and inverse problems.

\section{Numerical Algorithms}
\label{sec:numerics}
First, we summarize the numerical scheme for solving for forward and adjoint variables given $\vect{p}$, $\rho$ and $\kappa$. We also give the precise definition of how we choose the number and centers of the Gaussian basis functions $\{\phi_i(\vect{x})\}_{i=1}^m$, which we use to parameterize $c(\vect{x},0)$. Then, we discuss the optimization algorithm and how we impose the sparsity and ``$\max$'' constraints on $\boldsymbol{\phi}^T\vect{p}$.

\subsection{Forward solver numerics}\label{sec:fwd_num}
We use a pseudo-spectral Fourier method on a regular mesh for spatial discretization and employ 3D fast Fourier transforms to compute all spatial differential operators. We enforce our boundary conditions on the surface of the brain and CSF boundaries using a penalty approach (see~\cite{Gholami:2016a, Hogea:2008b}). We use periodic boundary conditions on $\Omega = [0, 2\pi]^3$.
We use an unconditionally stable, second order Strang operator-splitting method to numerically solve the forward and adjoint reaction-diffusion PDEs in Eq.~\eqref{eq:first_order_opt}. The forward equation is split into reaction and diffusion terms (for more details, see~\cite{Gholami:2016a, Hogea:2007b}). We use the Crank-Nicolson~\cite{crank1996} method for the diffusion term. This is an implicit method and the linear system is solved with a matrix-free preconditioned Conjugate Gradient method. We precondition the resulting implicit system of equations by solving them with constant coefficients computed as the average diffusion coefficient in the domain. The reaction term is solved analytically.  The main cost of the forward and adjoint solves are fast Fourier transforms to apply derivatives and preconditioners. 

\subsection{Selection of basis functions for the tumor initial condition}\label{sec:gaussians}
Given a tumor segmentation from an MRI image, we proceed to define the Gaussians, i.e., their number $m$, their centers $\{\vect{x}_i\}_{i=1}^m$, and the standard deviation $\sigma$ of each Gaussian.
To avoid spurious growth and diffusion of cancer cells in the CSF, we set the Gaussians to zero in this region. In other words, we have:
\begin{equation}
	\phi_i(\vect{x}) = \exp\left( -\frac{1}{2\sigma^2} \|\vect{x} - \vect{x}_i\|_2^2 \right) \mathbb{1}(m_{CSF}(\vect{x}) = 0),
\end{equation}
where $\mathbb{1}$ is the indicator function. We construct a candidate set of Gaussians whose locations are placed at a subset of a regular grid with spacing $\delta = 2 \sigma$ so that we can control the conditioning of  $M := \int_{\Omega} \boldsymbol{\phi} \boldsymbol{\phi}^T d\vect{x}$.\footnote{We empirically determine the ratio $\delta/\sigma$ by observing the condition number of $M$.} We set $\sigma = \lambda h_\mathrm{image}$, where $h_\mathrm{image}$ is the resolution of the input MRI image and $\lambda$ is an integer constant. Notice that $\sigma$ is a modeling parameter. If we refine the image resolution or the underlying discretization, $\sigma$ should remain unchanged.

\begin{figure}
	\centering
 	\includegraphics[width=0.35\textwidth]{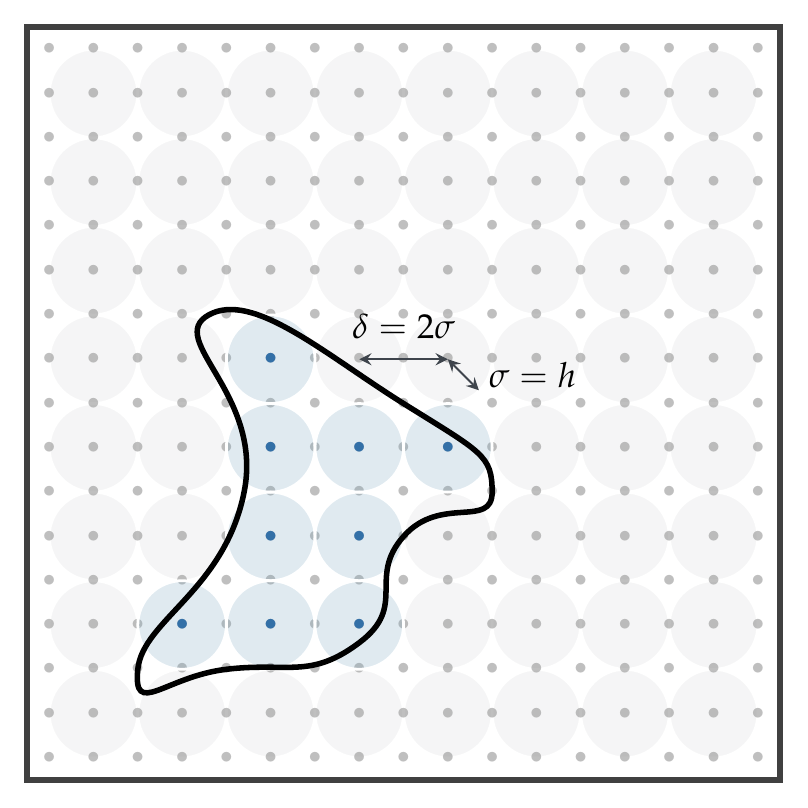}
 	\caption{\textit{A 2D illustration of the selection of Gaussian basis functions given the tumor segmentation boundary (figure modified from~\cite{Scheufele:2019b}). A candidate Gaussian set (light gray shaded circles) is constructed by seeding the entire domain with Gaussians with fixed variance $\sigma = h$ in this example. The Gaussians are centered at regular grid points $\vect{x}_i$ and are spaced at $\delta = 2\sigma$. If the volume fraction of tumor cells within any candidate Gaussian exceeds a user-defined threshold $\tau_{V}$, the candidate is selected (light blue shaded circles).}}\label{fig:gaussians}
\end{figure}

From this candidate set, we select the Gaussians based on the amount of tumor cells (volume fraction) within a radius of one standard deviation around the candidate location $\vect{x}_i $. We choose a candidate Gaussian if the volume of tumor around it exceeds a user-defined tolerance $\tau_{V}$, i.e.,
\begin{equation}
	\frac{\int_{B^\sigma(\vect{x}_i )} d(\vect{x})d\vect{x}}{\int_{B^\sigma(\vect{x}_i )} d\vect{x}} > \tau_V,
\end{equation}
where $B^\sigma(\vect{x}_i )$ is a ball of radius $\sigma$ around $\vect{x}_i$ and $d(\vect{x})$ is the tumor data. We show an illustration of this method in~\figref{fig:gaussians}.
This adaptive approach is fully automatic and is effective, for example, in identifying multifocal tumorous regions without any manual seeding.

\subsection{The $\ell_1$ regularized inverse problem}\label{sec:sparsity}
While an $L_2$ penalty is effective in regularizing the inverse problem for $\rho$, $\kappa$ and $\vect{p}$, it results in an initial condition that has an unrealistically large support and raises the question of how much we can trust the reconstructed $\rho$ and $\kappa$ coefficients. In fact, the reconstructed initial condition $c(\vect{x},0)$ closely resembles the data $d(\vect{x})$ up to a scaling, leading to $\kappa$ and $\rho$ being underestimated (see \secref{sec:AT} for numerical experiments highlighting this issue). As discussed in the introduction, to truly restrict $c(\vect{x},0)$, we need another modeling assumption -- that the tumor originated in a small number of locations. We call this property the sparsity of the initial condition.

\medskip \noindent \textbf{Enforcing sparsity. }The most common technique for imposing sparsity is to use the $\ell_1$ norm (see~\cite{Tropp10} for a comprehensive review of such techniques). Here, we assume that only a small number of Gaussians are activated and impose an $\ell_1$ penalty on the solution vector $\vect{p}$. We pursue an iterative recovery strategy similar to the algorithm described in~\cite{NEEDELL2009301}.
The main difficulty in using this algorithm directly is that our formulation involves a non-linear and non-convex inverse problem unlike the operators for the mismatch term in~\cite{NEEDELL2009301}. This also means we lose any theoretical guarantees of finding a solution. In particular, the crucial step of identifying the important components of the signal relies on certain properties of the sampling matrix, such as the restricted isometry property, which may no longer hold.
For our formulation, we use the gradient of the objective calculated using adjoints to identify the important components (similar to the original algorithm described in~\cite{NEEDELL2009301}). Before we outline our algorithm to reconstruct all the variables ($\vect{p}, \kappa$ and $\rho$), we describe our strategy to enforce the non-convex and non-differentiable max constraint in Eq.~\eqref{eq:constaint_p} which allows us to identify $\rho$.

\medskip \noindent \textbf{Enforcing the max constraint. }
While there are methods such as relaxing the max constraint to inequalities and using slack variables to handle them or mixed integer programming formulations, we do not pursue them here, in interest of algorithmic simplicity. Instead, we use the following strategy:
\bipa
	\item We start with an initial guess\footnote{We discuss our choice for this initial guess in~\secref{sec:setup}.}  for $\rho$ and invert for the parameters $(\vect{p}, \kappa)$. The resulting initial condition does not necessarily satisfy Eq.~\eqref{eq:constaint_p}.
	\item Then we re-scale the reconstructed initial condition appropriately to satisfy Eq.~\eqref{eq:constaint_p}. This fixes the scale of the reaction coefficient, as discussed in~\secref{sec:1d}.
	\item Finally, we invert for the variables $(\rho, \kappa)$ keeping the (re-scaled) initial condition fixed.
\eipa
\medskip

Before describing the complete algorithm, we introduce some notation. We write
	\begin{equation}
	\text{supp}(\vect{p}) = \{j: p_j \neq 0\}
	\end{equation}
for the \textit{support of $\vect{p}$}, and with
	\begin{equation}\label{eq:restriction}
	\vect{p}|^{s} = \begin{cases}
	p_j, & j \in S, \\
	0, &	\text{otherwise,}
	\end{cases}
	\end{equation}
  we denote the \textit{$s$ restriction of $\vect{p}$}, where $S$ is the set of indices of the $s$ largest-magnitude components of $\vect{p}$.

In a first phase, the algorithm simultaneously inverts for $\vect{p}$ and $\kappa$ following an iterative strategy:
\begin{enumerate}[(i)]
	\item \textit{Identify significant Gaussians: }Compute the gradient using the current approximation for $\vect{p}$ and $\kappa$. The largest components of the gradient of $\vect{p}$ indicate the location of important Gaussians and form the new support of $\vect{p}$ (cf.\;\algref{algo:cosamp_algo}, line\;\ref{alg:gaussian-identification}).
	\item \textit{Merge supports:} Merge the new support derived from the gradient with the support of the current approximation of $\vect{p}$ (cf.\;\algref{algo:cosamp_algo}, line\;\ref{alg:merge}).
	\item \textit{Signal estimation:} Solve an $L_2$ regularized inverse problem on the restricted subspace (defined by the merged support). This results in a new approximation for $\vect{p}$ and $\kappa$, (cf.\;\algref{algo:cosamp_algo}, line\;\ref{alg:estimate}).
	\item \textit{Signal restriction:} Restrict the new approximation to $\vect{p}$ to the required sparsity using Eq.~\eqref{eq:restriction}, (cf.\;\algref{algo:cosamp_algo}, line\;\ref{alg:restrict}).
  \item Iterate until stopping criterion is true. 
\end{enumerate}
We terminate the compressive sampling when a maximum number of iterations has been reached or the relative change in objective function value is below a user-defined threshold $\tau_J$. This stopping criterion is typically used in $\ell_1$ minimization problems. Other criteria such as monitoring the monotonic decrease of the mismatch two-norm have been used in subspace pursuit algorithms but are valid only in the linear case with restricted isometry properties. Hence, we do not employ these rules.

In a second phase, the signal estimation step is performed once more (cf.\;\algref{algo:cosamp_algo}, line\;\ref{alg:final_est}) in the subspace defined by the support of $\vect{p}$ to get the final approximation (rather than just keeping the largest entries). This step is important as the reconstructed initial condition is used in the subsequent inversion for $\rho$: We re-scale the initial condition and solve the tumor inversion problem for parameters $\rho$ and $\kappa$ keeping the initial condition fixed. Next, we describe the details of the signal estimation step.

\medskip
\noindent \textbf{The $L_2$ regularized inverse problem. }An algorithm for solving the $L_2$ regularized inverse problem (Eq.~\eqref{eq:tumor_opt}) for $\vect{p}$ and $\kappa$ was presented in~\cite{Gholami:2016a, Scheufele:2017a}. Here, we summarize the main components for solving the system of optimality conditions in Eq.~\eqref{eq:first_order_opt}. We employ a quasi-Newton method, L-BFGS~\cite[p.\,135ff]{Nocedal:2006a}) with matrix-free approximations to the inverse Hessian and the Mor{\'e}-Thuente criterion~\cite{More:1994a} for line-search. We enforce the bound constraints on our parameters $(\vect{p}, \kappa, \rho)$ using projected gradients to approximate the quasi-Newton Hessian. We terminate our solver when the relative change in the norm of the gradient falls below a user-defined threshold $\tau_g$ or when a maximal number of quasi-Newton iterations is reached. We discuss the choice for these parameters in \secref{sec:setup}. We compute our reference gradient with a zero initial guess for $(\vect{p}^0, \kappa^0)$.

We present a pseudocode in~\algref{algo:cosamp_algo}. 
\begin{algorithm}
	\caption{\textbf{: Compressive sampling for tumor inversion given sparsity level $s$ and initial guess for reaction coefficient $\rho^0$}}\label{algo:cosamp_algo}
	\begin{algorithmic}[1]
		\Procedure \Comment{\textit{Enforcing sparsity for the tumor inverse problem Eq.~\eqref{eq:tumor_opt}}}
		\State $\vect{p}^0 \gets 0$
		\State $\kappa^0 \gets 0$
		\State $i \gets 0$

		\While{stopping criterion is \textit{false}}
		\State $i \gets i + 1$
		\State $S \gets \text{supp}(\mathbf{g}_{\vect{p}}|^{2s}) $ \Comment{\textit{Identify significant Gaussians}\label{alg:gaussian-identification}}
		\State $S \gets S \cup \text{supp}(\vect{p}^{i - 1})$ \Comment{\textit{Merge supports}\label{alg:merge}}
		\State Invert for $(\vect{p}, \kappa)$ in subspace $S$ with fixed reaction, $\rho^0$  \Comment{\textit{Signal estimation using adjoints}\label{alg:estimate}}
		\State $\vect{p}^{i} \gets (\vect{p}^{i})|^{s}$ \Comment{\textit{Signal restriction to required sparsity}\label{alg:restrict}}
		\EndWhile\label{cosamp_loop}
		\State Invert for $(\vect{p}, \kappa)$ in $\text{supp}(\vect{p}^{i})$ with fixed reaction, $\rho^0$ \Comment{\textit{Signal estimation with final support}\label{alg:final_est}}
		\State $\vect{p}^{i} \gets \vect{p}^{i} / \max (\boldsymbol{\phi}^T \vect{p}^{i})$		\Comment{\textit{Initial condition re-scaling to satisfy Eq.~\eqref{eq:constaint_p}}}
		\State Invert for $(\rho, \kappa)$ with fixed initial condition  \Comment{\textit{Biophysical parameter inversion using adjoints}}
		\EndProcedure
	\end{algorithmic}
\end{algorithm}

\section{Results}
\label{sec:results}
We test the performance of our solver on synthetically generated tumors and clinical datasets.

\subsection{General setup} \label{sec:setup}
All 3D numerical experiments were executed on the Stampede2 system at the Texas Advanced Computing Center (TACC) at  The University of Texas at Austin on its dual socket Intel Xeon Platinum 8160 (``Skylake") nodes. Our solver is written in C++. We use MPI for parallelism compiled with Intel 17 compiler and 64 MPI tasks on 3 nodes for all our simulations. We use PETSc for all linear algebra operations, PETSc-TAO for the bound constrained L-BFGS optimization solver used in the $L_2$ regularized inverse problem, AccFFT for the 3D parallel FFTs and PnetCDF for I/O. We use a 2D pencil decomposition of our data for 3D distributed FFTs (see~\cite{Gholami:2016b} for the parallel implementation details). For 1D test-cases, we use \MATLAB R2018b.

\medskip \noindent \textbf{Performance measures.} The primary goal of our numerical experiments is to demonstrate the ability of the solver to reconstruct the reaction and diffusion coefficients, and to reconstruct the initial condition with good localization accuracy.
To evaluate the errors, we define the following error metrics to quantify our reconstruction:
\begin{enumerate}[(i)]
	\item relative error in $\kappa$ and  $\rho$ reconstruction:
	\begin{equation}
	e_{\kappa} = \frac{|\kappa^\text{rec} - \kappa^{\star}|}{|\kappa^{\star}|}, \quad
    e_{\rho} = \frac{|\rho^\text{rec} - \rho^{\star}|}{|\rho^{\star}|},
	\end{equation}
	where $\kappa^{\star}$ and $\rho^\star$ are the true coefficient values and $\kappa^\text{rec}$ and $\rho^\text{rec}$ are the reconstructed values,
    \item relative error in initial tumor reconstruction:
    \begin{equation}
    \elltwoIC =  \frac{\|c^{\text{rec}}(0) - c^\star(0)\|_{L_2(\Omega)}}{\|c^\star(0)\|_{L_2(\Omega)}},
    \end{equation}
    where $c^\text{rec}(0)$ is the reconstructed initial condition and $c^\star(0)$ is the ground truth initial condition,
	\item relative error in final tumor reconstruction:
	\begin{equation}
	\elltwoT = \frac{\|c^{\text{rec}}(1) - c^\star(1)\|_{L_2(\Omega)}}{\|c^\star(1)\|_{L_2(\Omega)}},
	\end{equation}
	where $c^{\text{rec}}(1)$ is the reconstructed tumor final concentration and $c^\star(1)$ is the final tumor concentration grown with the ground truth parameters $(\vect{p}^\star, \kappa^\star, \rho^\star)$,
	\item relative $\ell_1$ error in initial tumor parameterization:
	\begin{equation}
	\elltwoICP = \frac{  \| \vect{p}^\text{rec} - \vect{p}^\star \|_{\ell_1}   }{   \| \vect{p}^\star \|_{\ell_1} },
	\end{equation}
    with reconstructed parameterization vector $\vect{p}^\text{rec}$,
	\item relative change in the norm of the gradient to help assess the convergence of our numerical scheme:
	\begin{equation}
		\relG = \frac{\| g \|_{\ell_2}}{\| g^0 \|_{\ell_2}},
	\end{equation}
	where $g$ is the gradient at the final iteration and $g^0$ is the gradient with respect to the initial solution guess.
\end{enumerate}
We evaluate the performance of two solvers, namely the \textit{$\mathbold{L_2}$} solver and the \textbf{\textit{CS}} solver. The former refers to the $L_2$ (two-norm) regularized solver introduced in our previous work~\cite{Gholami:2016a, Scheufele:2017a} that does not use any sparsity constraints, while the latter is the compressive sampling solver described in~\algref{algo:cosamp_algo}.

\begin{table}
	\centering
	\caption[Numerical parameters]{\label{tab:num_params}Numerical parameters and their values used for tumor inversion.}
	\setlength\tabcolsep{10pt}\small
	\begin{tabular}{lc}
		\toprule
		Parameter              & Value            \\
		\midrule
		\textit{spatial discretization for synthetic tumors}      &$128^3$    \\
\highlightrow		\textit{volume fraction for Gaussian basis selection}, $\tau_{V}$ & $0.99$ \\
        \textit{standard deviation of Gaussian basis functions}, $\sigma$ & $2\pi/64$ \\
\highlightrow		\textit{relative gradient tolerance}, $\tau_g$		&  $\num{1E-5}$ \\
		\textit{relative objective tolerance}, $\tau_J$		&  $\num{1E-5}$ \\
\highlightrow		\textit{maximum quasi-Newton iterations}		&	$50$ \\
		\textit{maximum compressive sampling iterations}	& $2$ \\
\highlightrow		\textit{sparsity level}, $s$		&	10	\\
		\textit{regularization parameter}, $\beta$ &	$\num{1E-4}$ \\
\highlightrow		\textit{initial guess for reaction coefficient}, $\rho^0$ & 	$\num{15}$ \\
		\bottomrule
	\end{tabular}
\end{table}

\medskip \noindent \textbf{Numerical parameters:} We list all numerical parameters in \tabref{tab:num_params} and outline any parameter change specific to a test-case in its respective section. We now provide a justification for our parameter choices:
\begin{enumerate}[(i)]
	\item For the selection of Gaussians, we choose a volume fraction $\tau_{V}$ of $\num{0.99}$ to ensure that only Gaussians occupied completely by tumorous cells within a neighbourhood of one standard deviation $\sigma$ get selected. We consider Gaussians with smaller volume fractions as unimportant candidates for the support of the tumor initial condition and hence we neglect them.
	\item We refer to \cite{Scheufele:2017a} for the $L_2$ tumor regularization parameter $\beta$, which was chosen based on the data reconstruction quality and an L-curve analysis. For the compressive sampling (\textbf{\textit{CS}}) solver, we set $\beta$ to zero as the inversion is performed for typically 10--20 parameters and we have experimentally found that no regularization is necessary.
	\item We pick the relative gradient convergence tolerance $\tau_g$ for all $L_2$ solves (Eq.~\eqref{eq:tumor_opt}) to be $\num{1E-5}$ in order to get sufficient accuracy in the reconstruction.\footnote{We note that our formulation is based on an optimize-then-discretize approach and hence our gradient is only accurate up to discretization errors.}
	\item We set a maximal number of quasi-Newton iterations as 50 to prevent prohibitively large times-to-solution. But in most of our experiments, the target tolerance is reached within the given iteration budget. In the rare occasion this fails to happen (for example, in test-cases with extremely large tumors), we observe that additional quasi-Newton iterations do not lead to significant accuracy gains, but they do significantly increase the computational cost.
	\item Based on numerical experimentation with different values of reaction and diffusion coefficients, we find $\rho = 15$ to be a good representative reaction coefficient to capture quite ``large" tumors ({\raise.17ex\hbox{$\scriptstyle\sim$}}30\% of white matter volume). Hence, we set this to be our initial guess for the reaction coefficient ($\rho^0$).
	\item The required sparsity level is a question that needs to be addressed empirically using a large study using a clinical dataset. Here, we hypothesize 5--10 activated sites for tumor initiation and verify numerically that, using this sparsity, we capture important activations in all our synthetic test-cases. For clinical studies, we plan to use an image-analysis based tool to estimate the focality of the tumor and then assign a number of sites per focal region. This is beyond the scope of this paper.
	\item In all our experiments, we observe inconsequential changes in sparsity after two compressive sampling iterations and hence set it as an upper bound to the number of iterations for the $\ell_1$ sparsity.
\end{enumerate}

We examine different types of test-cases to illustrate the performance of our model. We describe the setup of each experiment and the corresponding results in the following sub-sections.

\subsection{1D constant coefficients test-case}

We start with a simple 1D example to demonstrate that the \textbf{\textit{CS}} solver outperforms the \textit{$\mathbold{L_2}$} solver.

\medskip \noindent \textit{Setup. }The data $d(\vect{x})$ is generated using the non-linear forward model with constant reaction and diffusion coefficients. We use a spatial resolution of $256$ grid points in $\Omega = [0,2\pi]$ and a temporal resolution of 100 time steps within a time horizon of 1.
We consider two variations for our test-cases: \\

\begin{tabular}{lllll}
\textit{(i)}   &\textit{1D-C1: } &  mono-focal tumor  & $\rho^{\star}=10$ & $\kappa^{\star}=0.05$   \\
\textit{(ii)}  &\textit{1D-C2: } &  multi-focal tumor & $\rho^{\star}=10$ & $\kappa^{\star}=0.05$   \\
\end{tabular} \\

\begin{figure}
    \centering
    \subfloat[\textbf{\textit{1D-C1}\label{fig:1d_a}} : \textit{$\mathbold{L_2}$ \textbf{solver}}]{\includegraphics[width=0.5\textwidth]{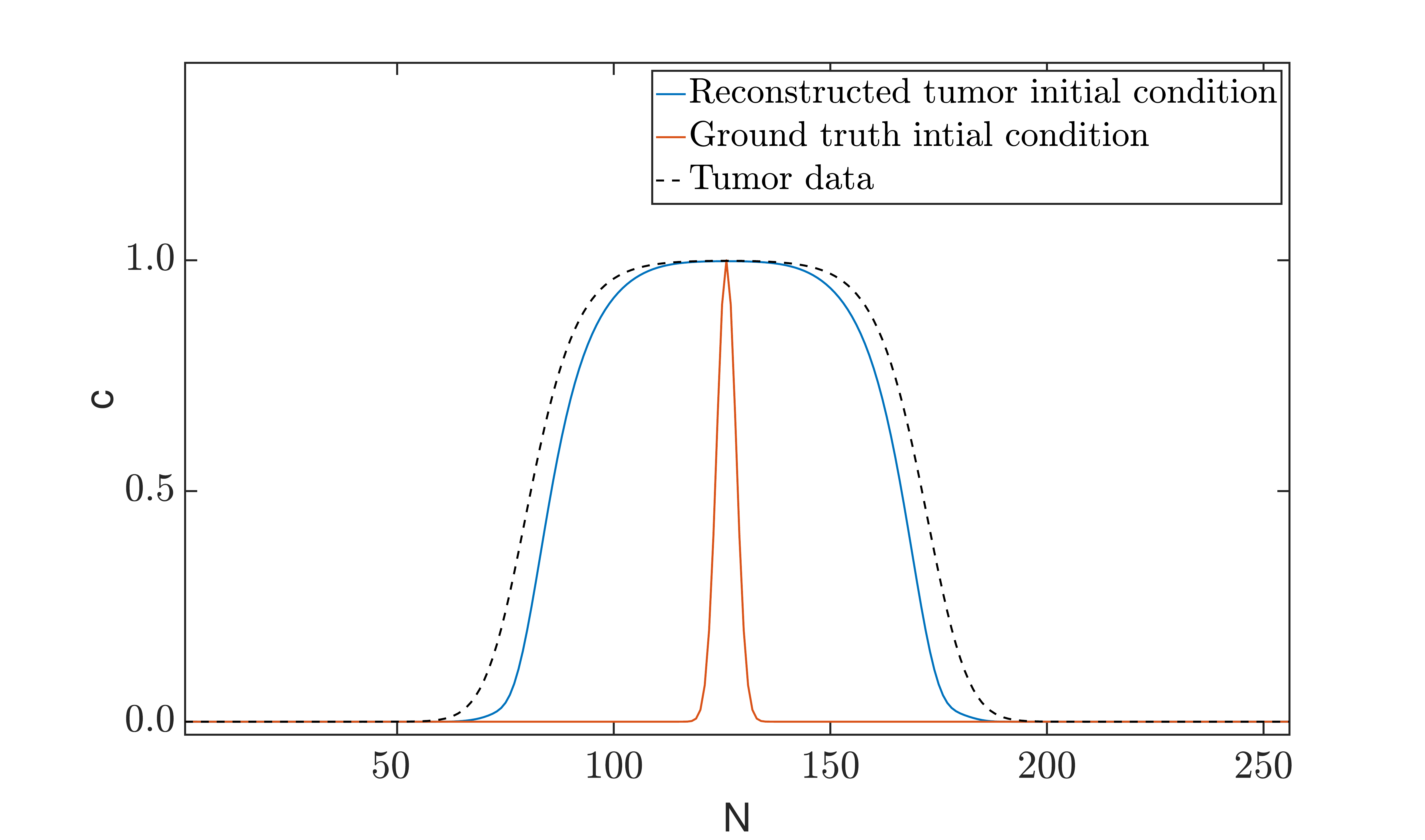}}%
    \hfill
    \subfloat[\textbf{\textit{1D-C1}\label{fig:1d_b}} : \textbf{\textit{CS solver}}]{\includegraphics[width=0.5\textwidth]{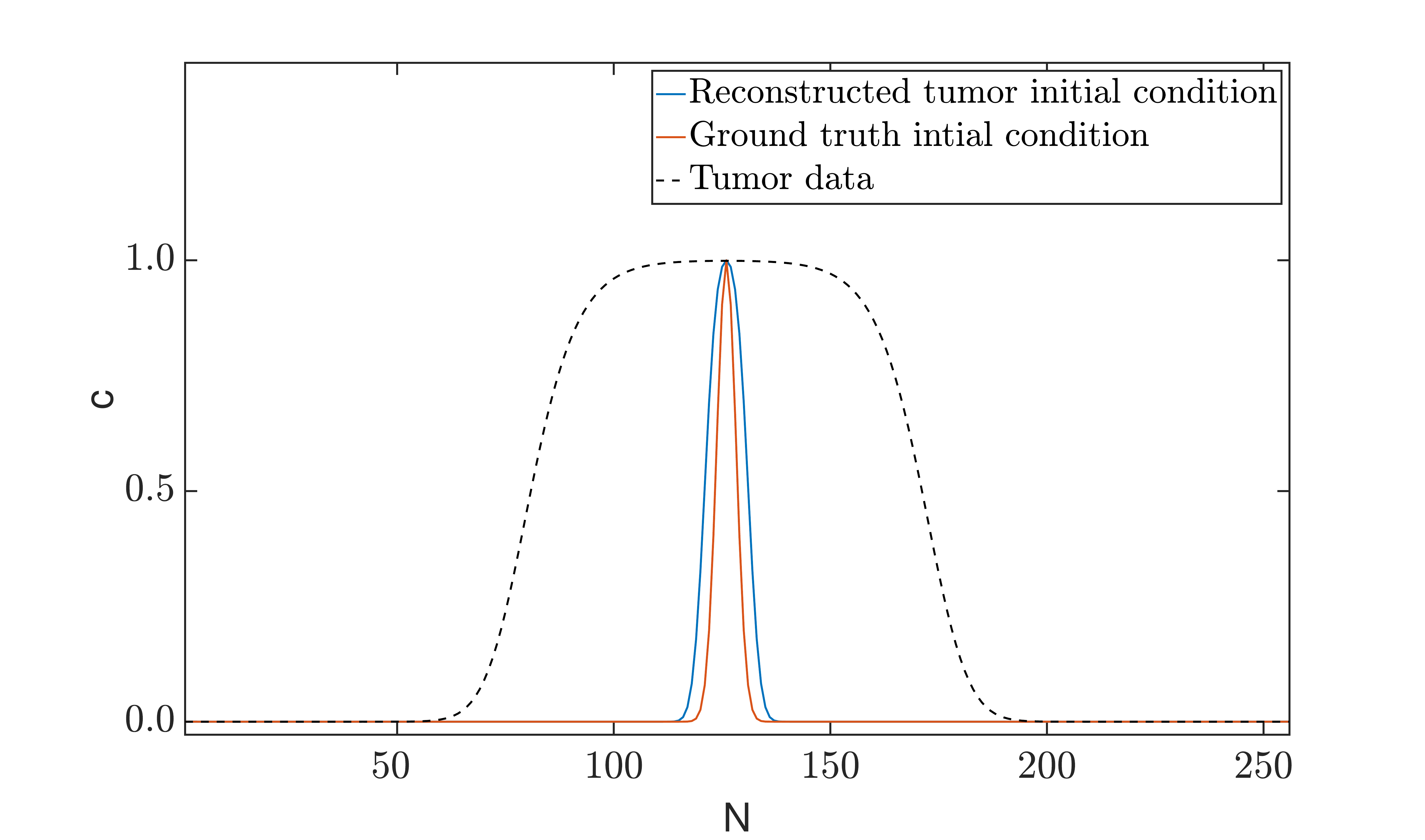}}
    \hfill
    \subfloat[\textbf{\textit{1D-C2}\label{fig:1d_c}} : \textit{$\mathbold{L_2}$ \textbf{solver}}]{\includegraphics[width=0.5\textwidth]{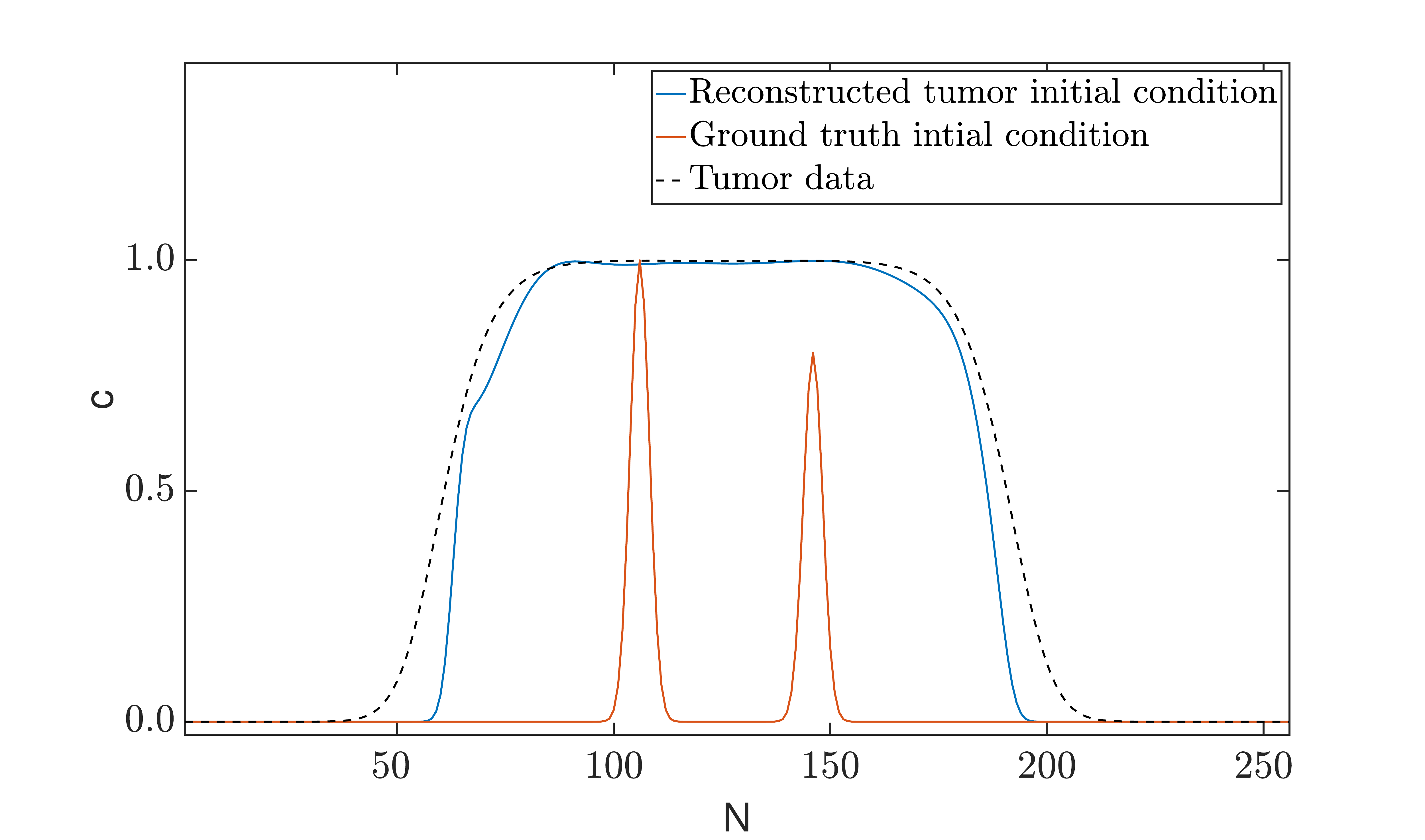}}%
    \hfill
    \subfloat[\textbf{\textit{1D-C2}\label{fig:1d_d}} : \textbf{\textit{CS solver}}]{\includegraphics[width=0.5\textwidth]{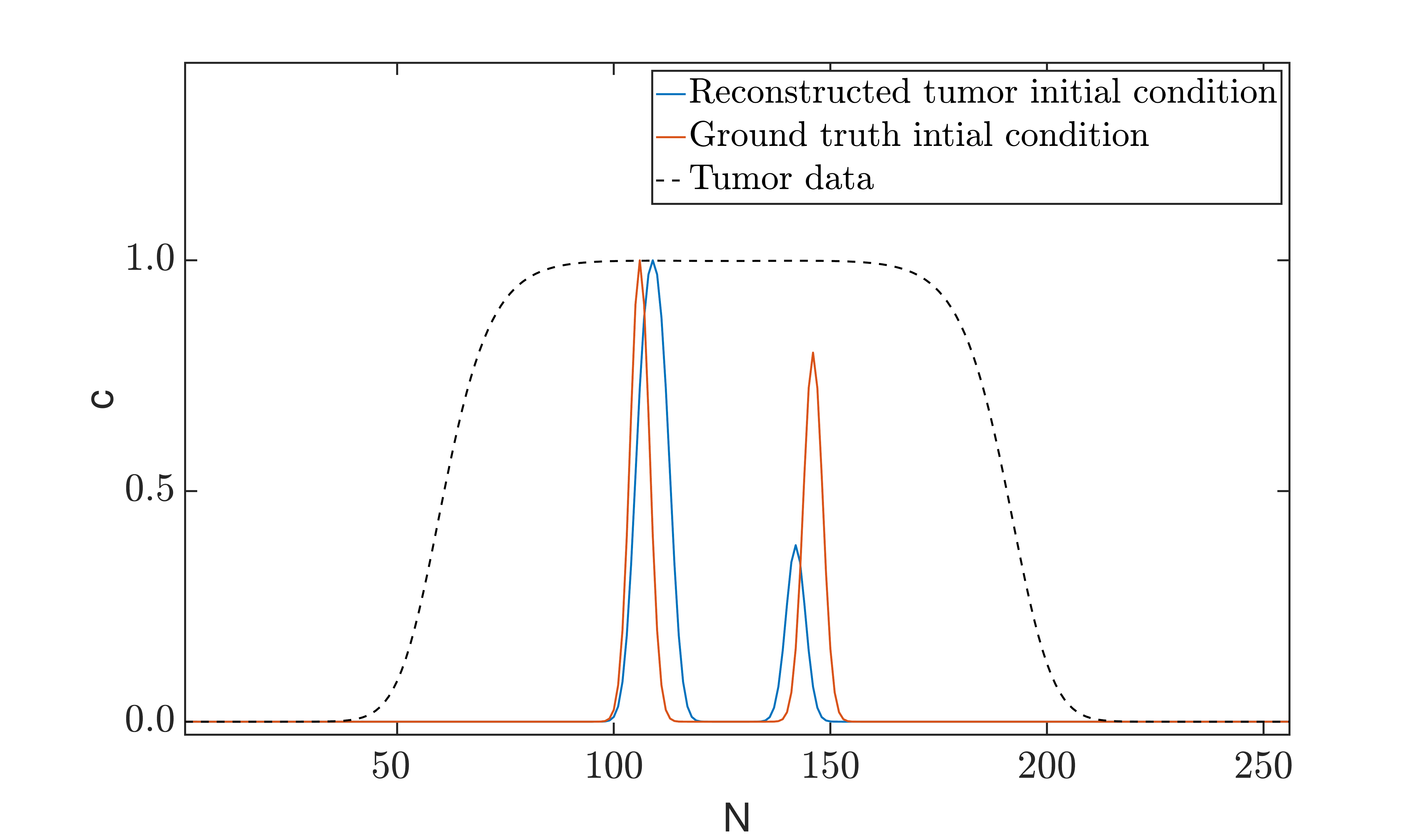}}
    \hfill
    \caption{\textit{Qualitative results for \textbf{1D constant coefficients} test-case: The images show the tumor initial condition reconstruction using the two approaches. The \textit{$\mathbold{L_2}$} solver predicts a complex initial condition and hence fails to reconstruct the biophysical parameters accurately. Enforcing sparsity helps in localizing the initial condition and hence enjoys significant performance improvements in parameter reconstruction. Despite these differences, we note that both solvers provide excellent reconstruction of the tumor data (see \tabref{tab:1d_tc} for the performance measures).}\label{fig:1d}}
\end{figure}

\noindent We consider more variations for our 3D test-cases to capture the heterogeneity in tumor shapes generally observed in clinical data. We report our performance measures, relative gradients and solver timings in \tabref{tab:1d_tc} and present qualitative results of the reconstruction in \figref{fig:1d}.

\medskip \noindent \textit{Observations. }While both solvers are able to fit the data satisfactorily (reconstruction errors less than $5$\%), we observe that the compressive sampling (\textbf{\textit{CS}}) solver outperforms its \textit{$\mathbold{L_2}$} counterpart in biophysical parameter estimation. This is evident in~\figref{fig:1d_a} and~\figref{fig:1d_c}, where the reconstructed initial conditions resemble grown tumors compared to the sparse ground truth. This can also be observed in the $\ell_1$ relative error norms of the initial condition parameterization ($\elltwoICP$) which show about an order of magnitude difference between the two solvers. Here, we would like to emphasize that it is because of the unknown reaction coefficient that the \textit{$\mathbold{L_2}$} solver reconstructs a rich initial condition, and consequently a low (and inaccurate) reaction coefficient. Hence, the tumor does not grow much, and instead diffuses to fit the data -- this is seen in the reconstructed values for the reaction and diffusion coefficients. As we shall see, this issue is mitigated a little in 3D, but still persists.
We also note that the initial conditions might not be perfectly recoverable due to the ill-conditioned nature of the inverse problem. For instance, the multi-focal test-case (\textit{1D-C2}) finds activations in neighbouring Gaussians to the ground truth. This localization problem is unavoidable due to ill-conditioning. If the time horizon is so large that the maximum tumor concentration is everywhere in the spatial domain, it is impossible to recover any information about the initial location of the tumor. So, the larger the tumor (or equivalently, the larger the time horizon),  the harder it is to localize the initial condition.

{
	\setlength\tabcolsep{6pt}
	\begin{table}
		\caption{\textit{
			Quantitative results for the \textbf{1D constant coefficients} test-case using the two-norm regularized solver (\textit{$\mathbold{L_2}$}) and the compressive sampling (\textbf{\textit{CS}}) solver. The ground truth forward solver parameters are indicated by '$^\star$' and their respective reconstructed coefficient values by '$\text{rec}$'. We use a spatial discretization of  $2\pi/256$ with 256 grid points and a time discretization of 0.01 with 100 time steps. We consider the ground truth for the model constants as $\rho^\star = 10, \kappa^\star = 0.05$ to ensure a large tumor (larger tumors correspond to larger time horizons which make inverting for initial conditions hard). We report the relative error in inverted coefficient values ($e_{\kappa}, e_{\rho}$), the relative error in final tumor reconstruction ($\elltwoT$), the relative error in the reconstructed initial condition ($\elltwoIC$), the relative error in the parameterization of the initial condition ($\elltwoICP$), the relative change in the gradient ($\relG$) and solver timings ($T_{inv}$) in seconds.}
			\label{tab:1d_tc}
		}
		\centering
		\begin{scriptsize}
			\makebox[\textwidth]{\centering
				\begin{tabular}{cccc|cLcLcLLcc}
					\toprule
					Test case & Solver & $\kappa^\star$ &  $\rho^\star$ & $\kappa^\text{rec}$ &  $e_{\kappa}$  & $\rho^\text{rec}$ & $e_{\rho}$ & $\elltwoT$ & $\elltwoIC$  &  $\elltwoICP$ & $\relG$ & $T_{inv}$\,[s]  \\
					\midrule

					\multirow{2}{*}{{\scriptsize \centering\textbf{\textit{1D-C1}}}}  &  \textit{$\mathbold{L_2}$} &  \num{5E-2} &	\num{1E1} & \num{1.026033e-02}   & \num{7.948E-1}	 & \num{1.05E0}	& \num{8.688E-1}	 & 	\num{1.220493e-03}& \num{3.951e+00} & \num{1.431e+01}	&  \num{6.690639e-04}   &	\num{7.016E0}	\\

					&  \textbf{\textit{CS}}  &  \num{5E-2} &	\num{1E1} &  \num{4.761684e-02}   & \num{4.77E-2}	& \num{9.4106E0}	& \num{5.89E-2}	 & 	\num{1.909789e-04 }& \num{6.486e-01} & \num{2.098e+00}	& \num{3.135666e-06}  &	\num{9.053E0}	\\

					\midrule

					\multirow{2}{*}{{\scriptsize \centering\textbf{\textit{1D-C2}}}}  &  \textit{$\mathbold{L_2}$} &  \num{5E-2} &	\num{1E1} &  \num{1.708996e-02}   & \num{6.582E-1}	& \num{1.2E0}	& \num{8.5E-1}	 & 	\num{1.571325e-03} & \num{3.840e+00} & \num{1.186e+01}	&  \num{4.339336e-05 }   &	\num{8.13E0}	\\

					 &   \textbf{\textit{CS}} &  \num{5E-2}  &	\num{1E1} & \num{5.066554e-02}   & \num{1.33E-2}	& \num{1.09412E1}	& \num{9.41E-2}	 & 	\num{5.093427e-02}& \num{8.803e-01}	& \num{1.638e+00}&  \num{2.939852e-06 }   &	\num{1.1724E1}	\\

					\bottomrule
			\end{tabular}}
		\end{scriptsize}
	\end{table}
}

\begin{figure}
	\centering
    \subfloat[]{\includegraphics[width=0.8\textwidth]{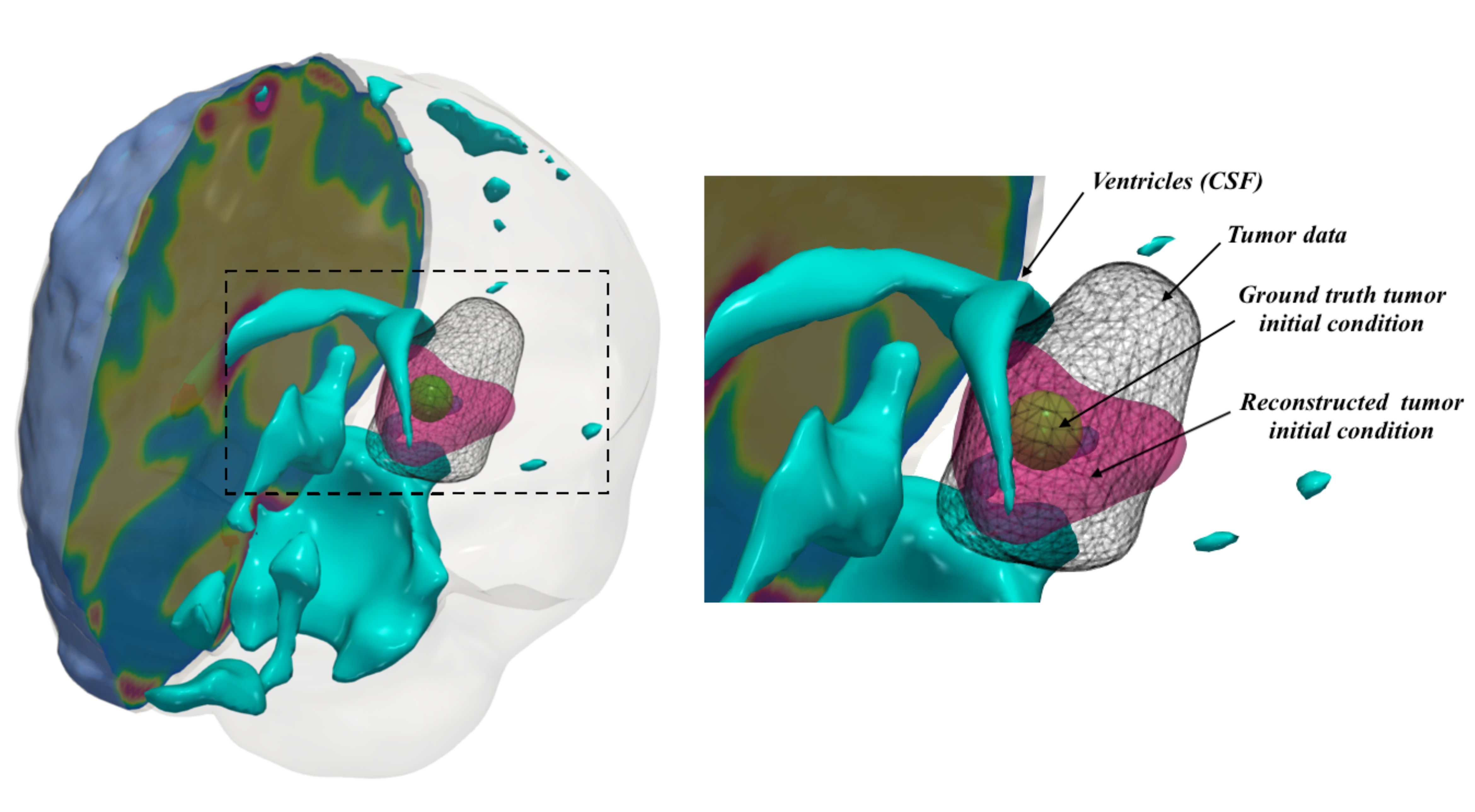}}\\%
	\subfloat[\textit{$\mathbold{L_2}$ }]{\includegraphics[width=0.45\textwidth]{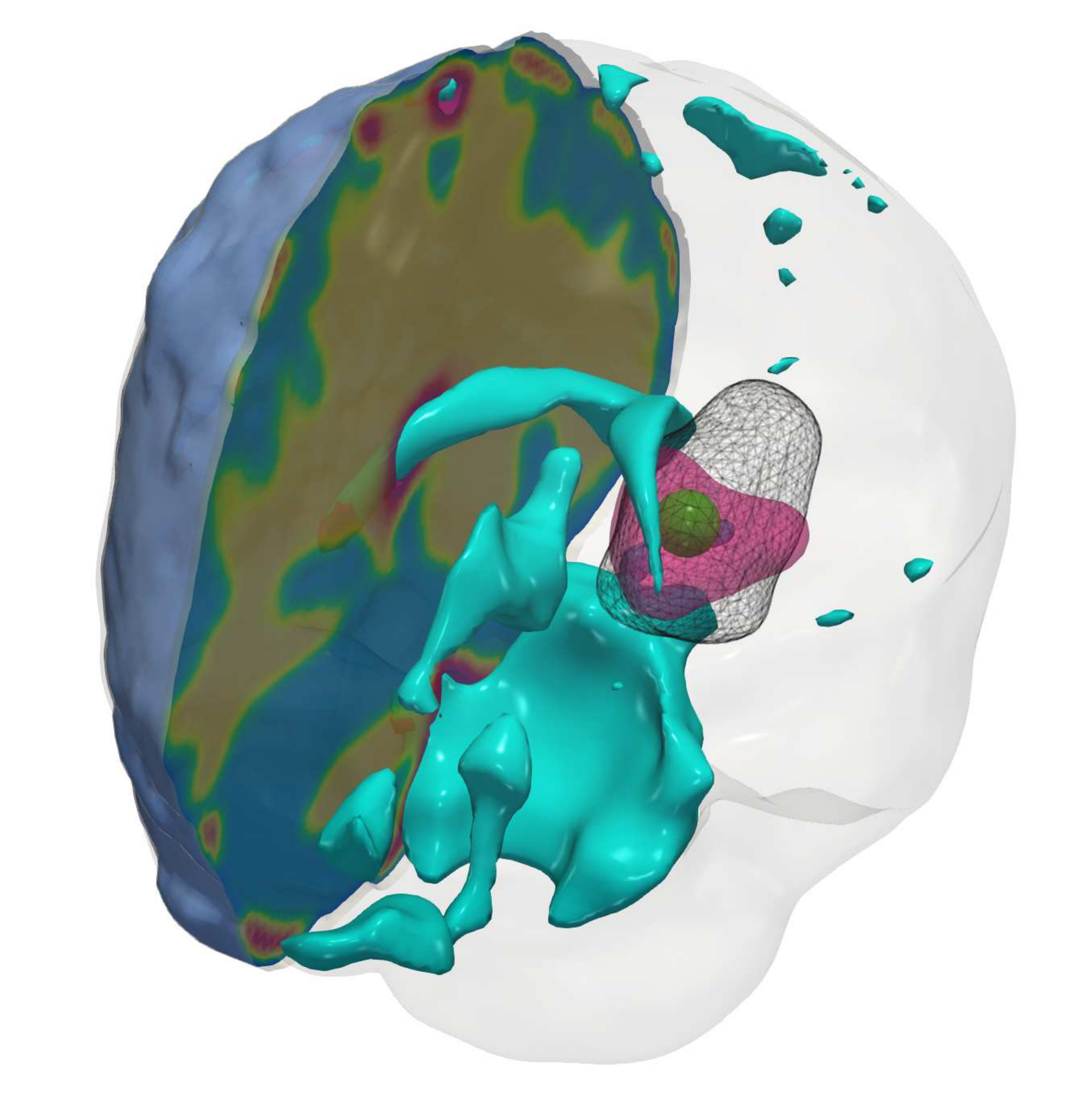}}%
	\subfloat[\textbf{\textit{CS}}]{\includegraphics[width=0.45\textwidth]{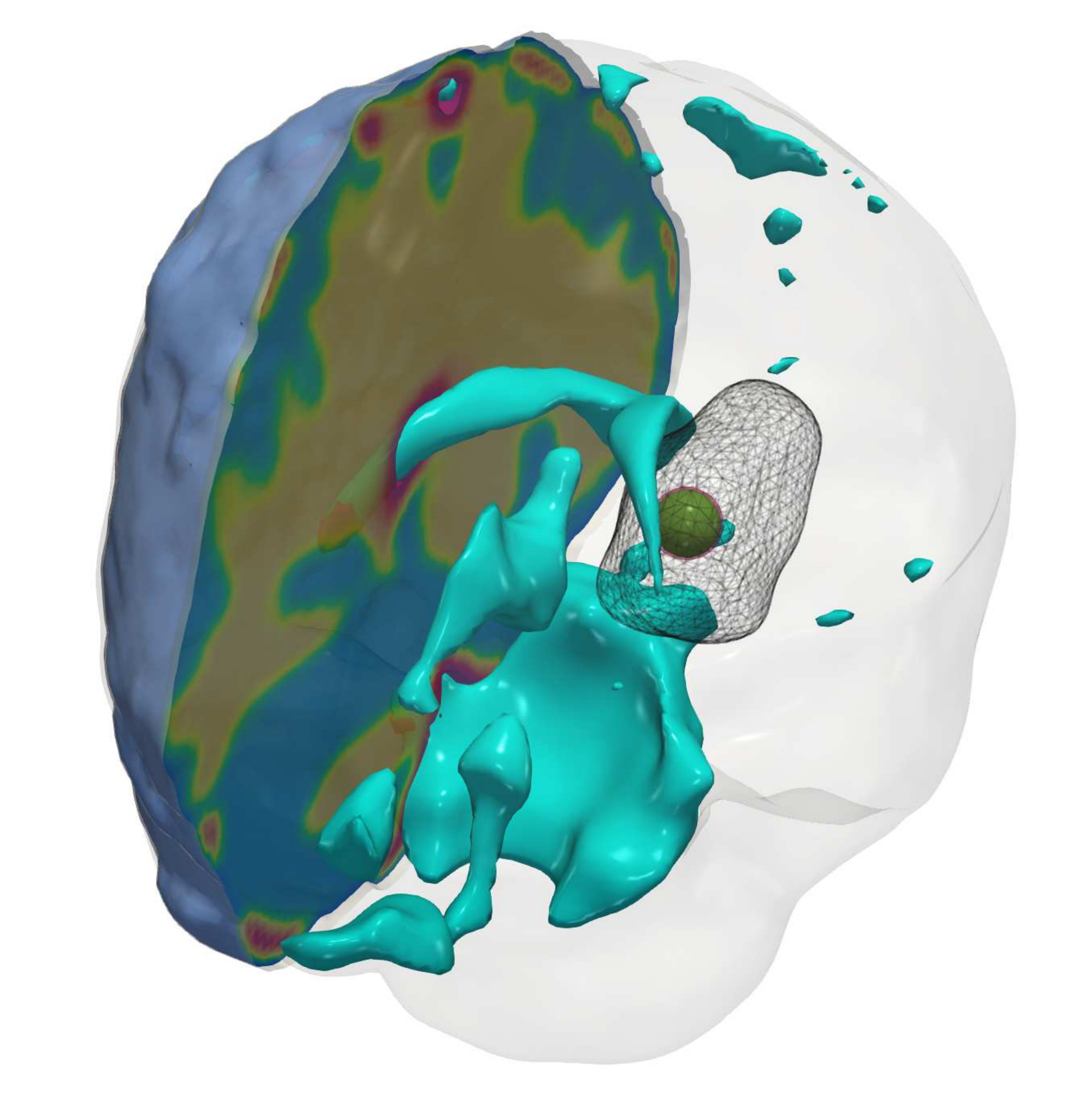}}
	\caption{\textit{(a) Description of the visualized brain sub-structures: The figure on the left shows an isometric view of the brain with different tumorous and healthy tissue sub-structures. The figure on the right zooms in on the tumor region and describes each structure. (b, c) Qualitative results for the \textbf{artificial tumor} test-case \textbf{\textit{AT-C1}} (a mono-focal, mostly proliferative tumor) with growth parameters $\rho^\star = 8,~\kappa^\star = 0.025$.  The images show the tumor data (gray wireframe), ground truth tumor initial condition (green volume), reconstructed tumor initial condition (magenta volume), ventricles (cyan volume), and a section of the patient brain geometry. We observe exact reconstruction for smaller tumors with sparsity constraints. The \textit{$\mathbold{L_2}$} solver fails to obtain the same reconstruction due to its inability to determine the correct reaction coefficient. We refer the reader to~\tabref{tab:synthetic_tc} for the quality of final tumor reconstruction.}\label{fig:at_c1}}
\end{figure}
\begin{figure}
	\subfloat[\textit{$\mathbold{L_2}$ }]{\includegraphics[width=0.45\textwidth]{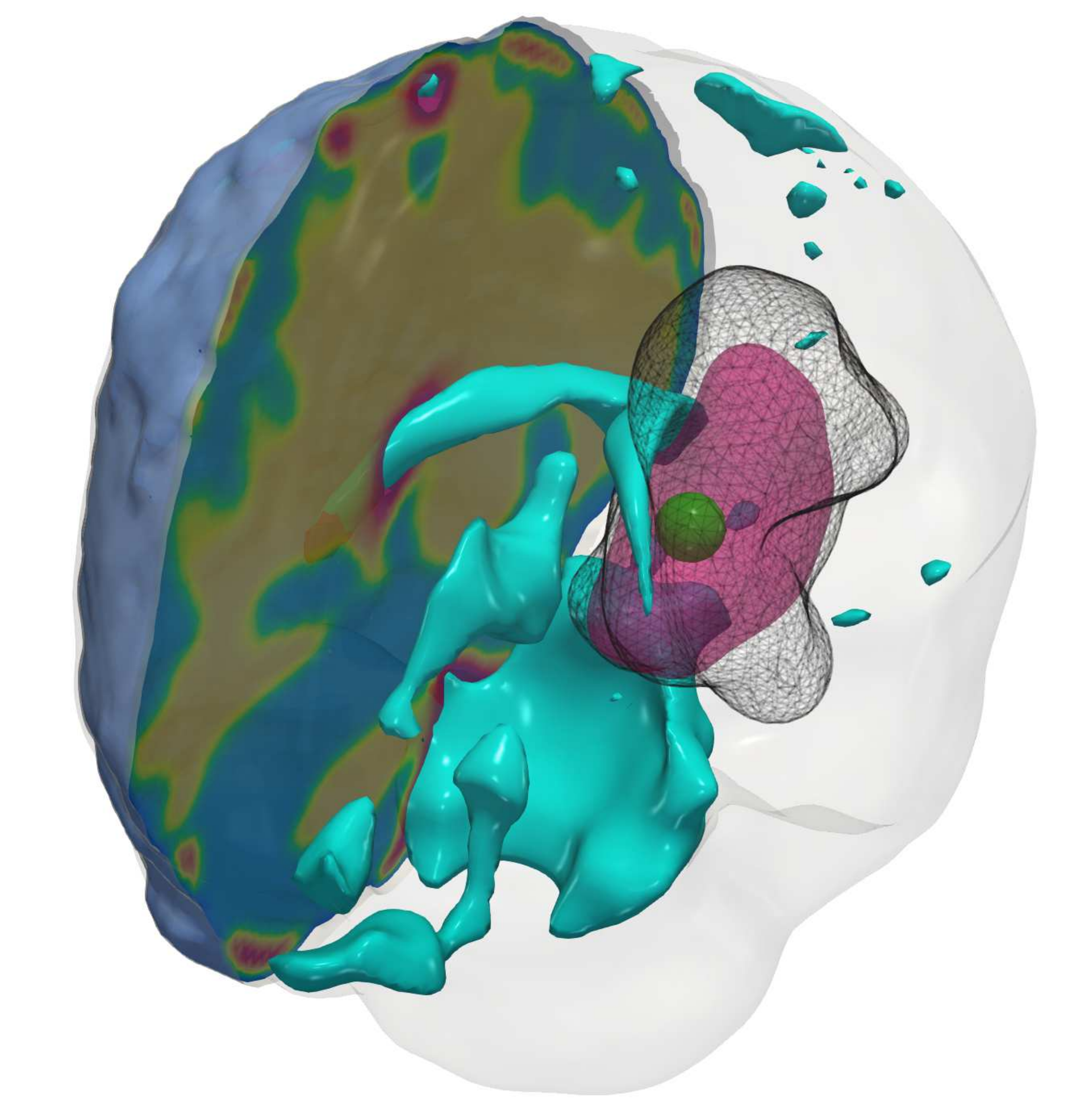}}%
	\subfloat[\textbf{\textit{CS}}]{\includegraphics[width=0.45\textwidth]{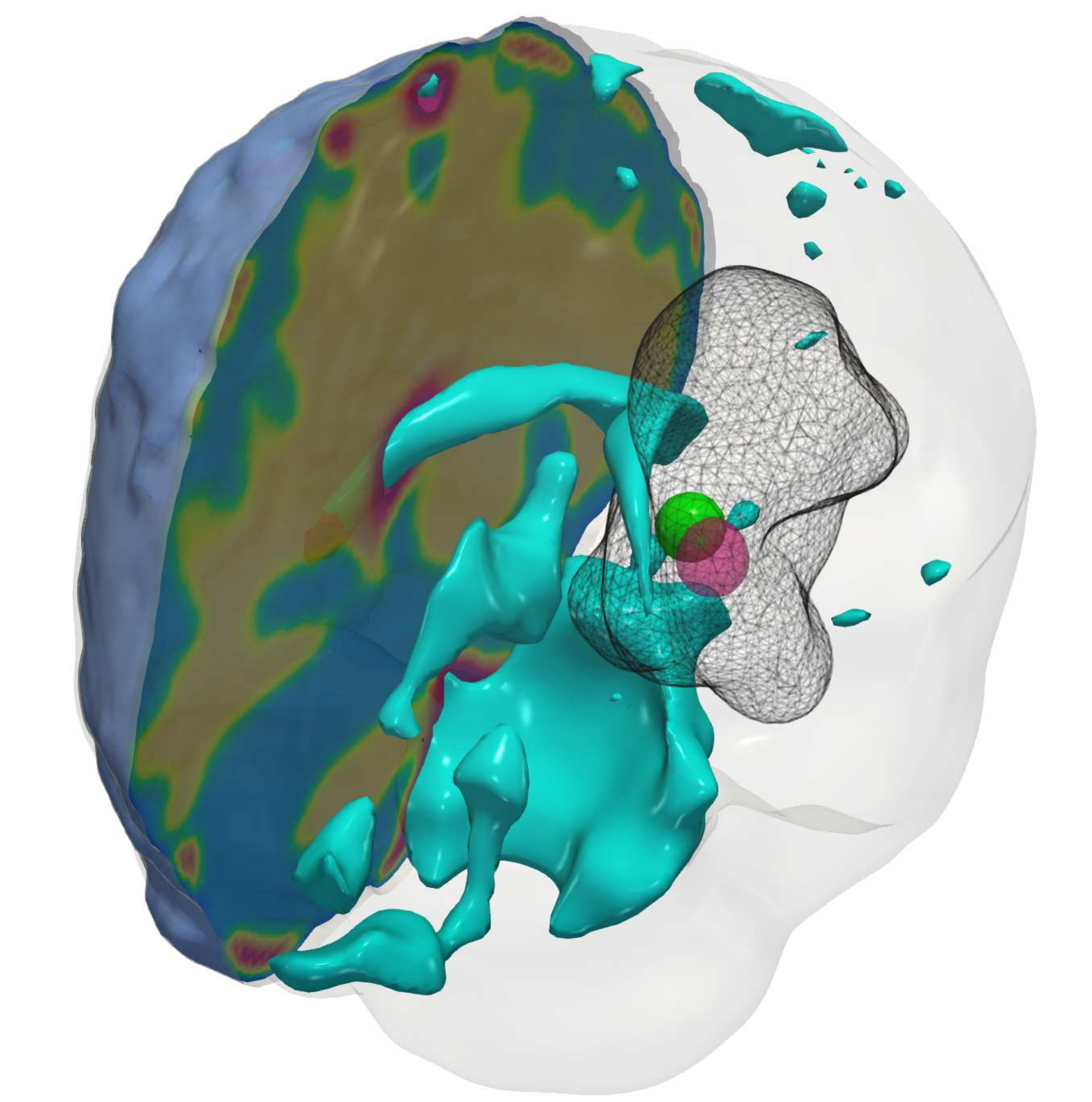}}
	\caption{\textit{Qualitative results for the \textbf{artificial tumor} test-case \textbf{\textit{AT-C2}} (a mono-focal tumor that has grown more and is infiltrative) with growth parameters $\rho^\star = 12,~\kappa^\star = 0.05$. The images show the tumor data (gray wireframe), ground truth tumor initial condition (green volume), reconstructed tumor initial condition (magenta volume), ventricles (cyan volume), and a section of the patient brain geometry. As with smaller tumors, the \textit{$\mathbold{L_2}$ solver predicts an initial condition reconstruction that has a large support and lacks localization capabilities. Hence, the solver is unable to determine the reaction coefficient accurately. While sparsity constraints help in localizing the initial condition, an exact initial condition reconstruction is difficult due to the ill-conditioned inverse problem. }}\label{fig:at_c2}}
\end{figure}
\begin{figure}
	\centering
	\subfloat[\textit{$\mathbold{L_2}$} ]{\includegraphics[width=0.45\textwidth]{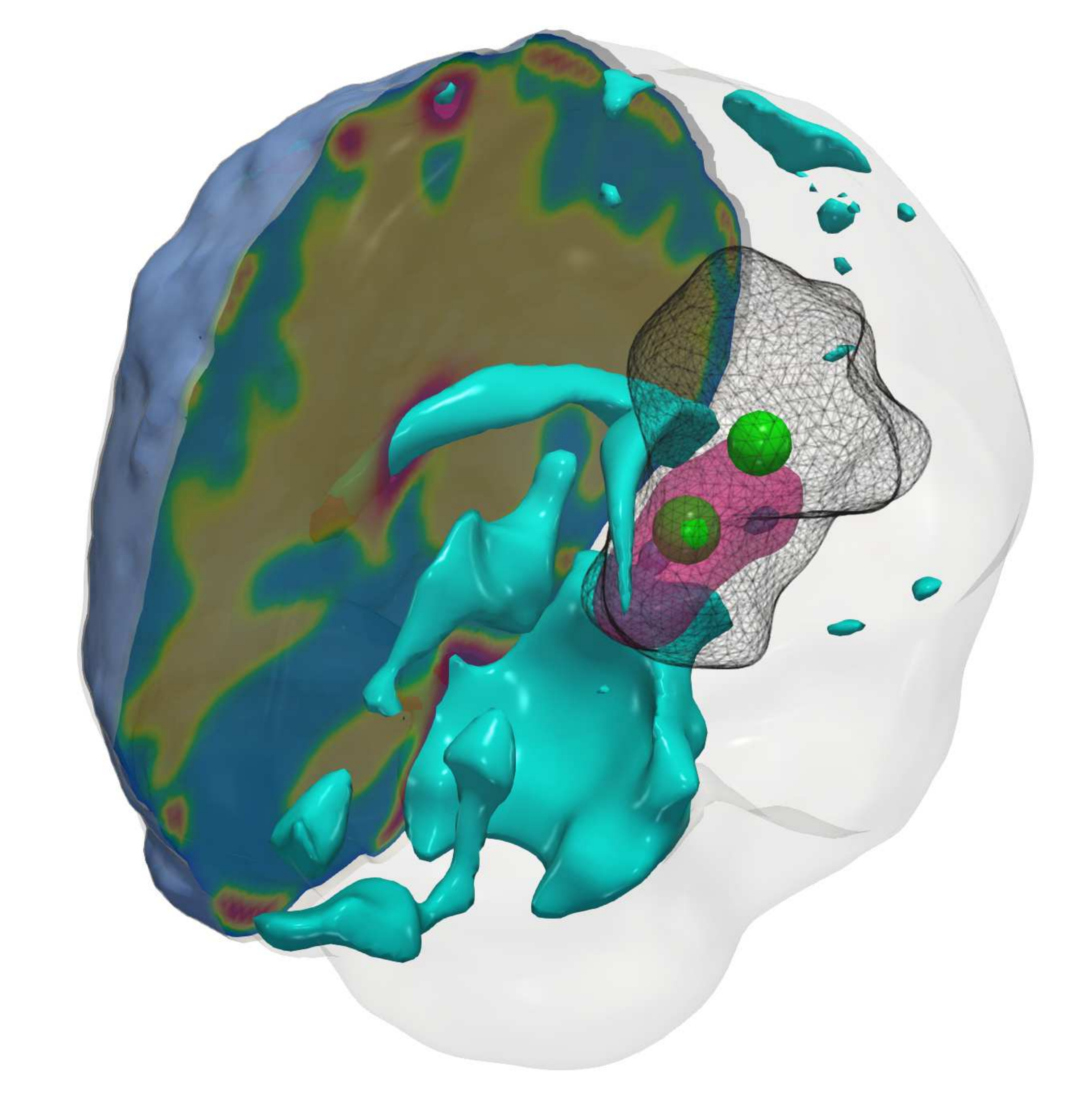}}%
	\subfloat[\textbf{\textit{CS}}]{\includegraphics[width=0.45\textwidth]{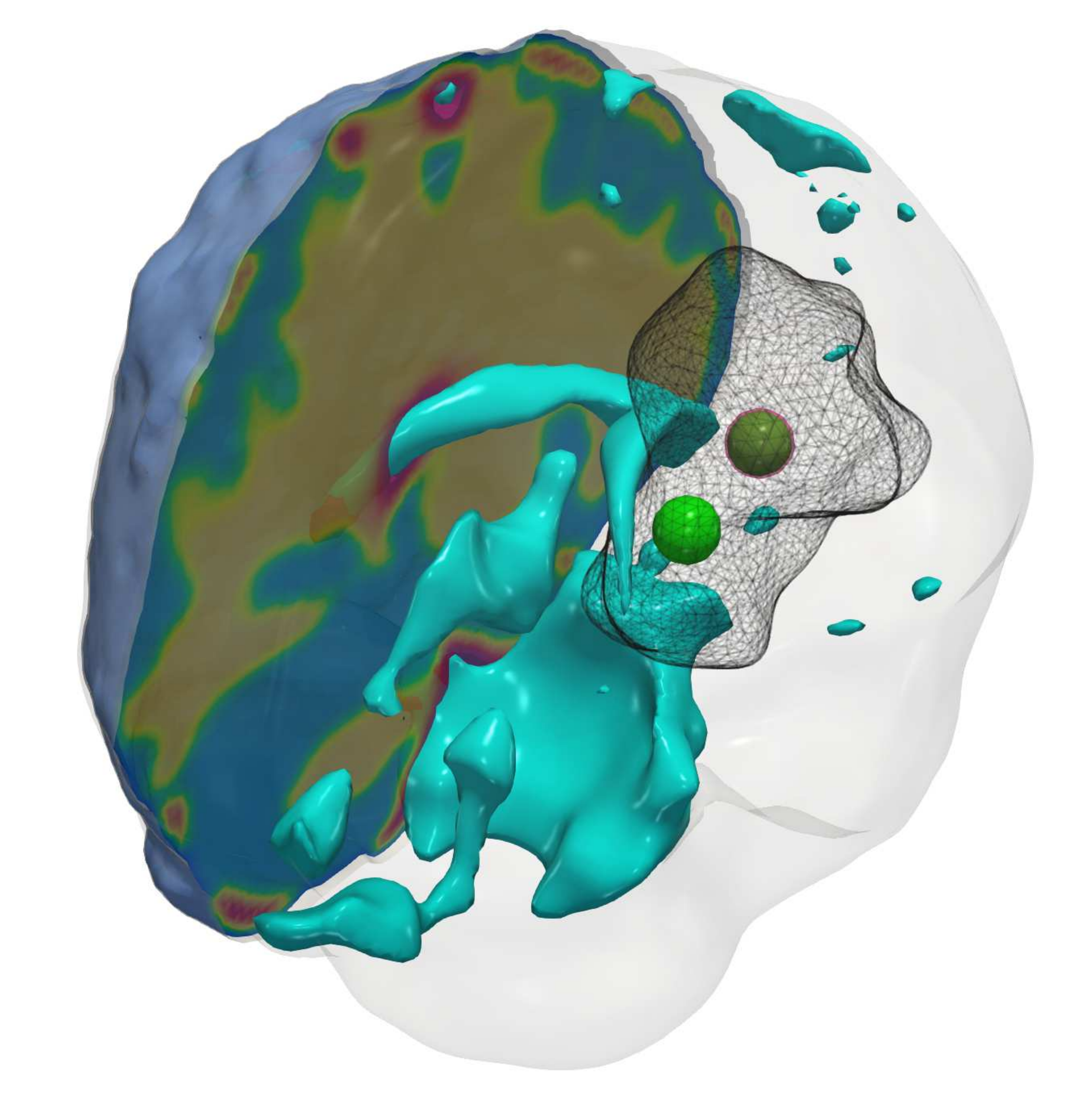}}
	\caption{\textit{Qualitative results for the \textbf{artificial tumor} test-case \textbf{\textit{AT-C3}} (a multi-focal tumor) with growth parameters $\rho^\star = 10,~\kappa^\star = 0.025$.  The images show the tumor data (gray wireframe), ground truth tumor initial condition (green volume), reconstructed tumor initial condition (magenta volume), ventricles (cyan volume), and a section of the patient brain geometry. The \textbf{\textit{CS}} solver activates one of the Gaussians exactly, while assigning smaller activations (not pictured) around the second ground truth Gaussian (similar to the 1D analogous test-case \textit{1D-C2}). }\label{fig:at_c3}}
\end{figure}
\begin{figure}
	\subfloat[\textit{$\mathbold{L_2}$ }]{\includegraphics[width=0.45\textwidth]{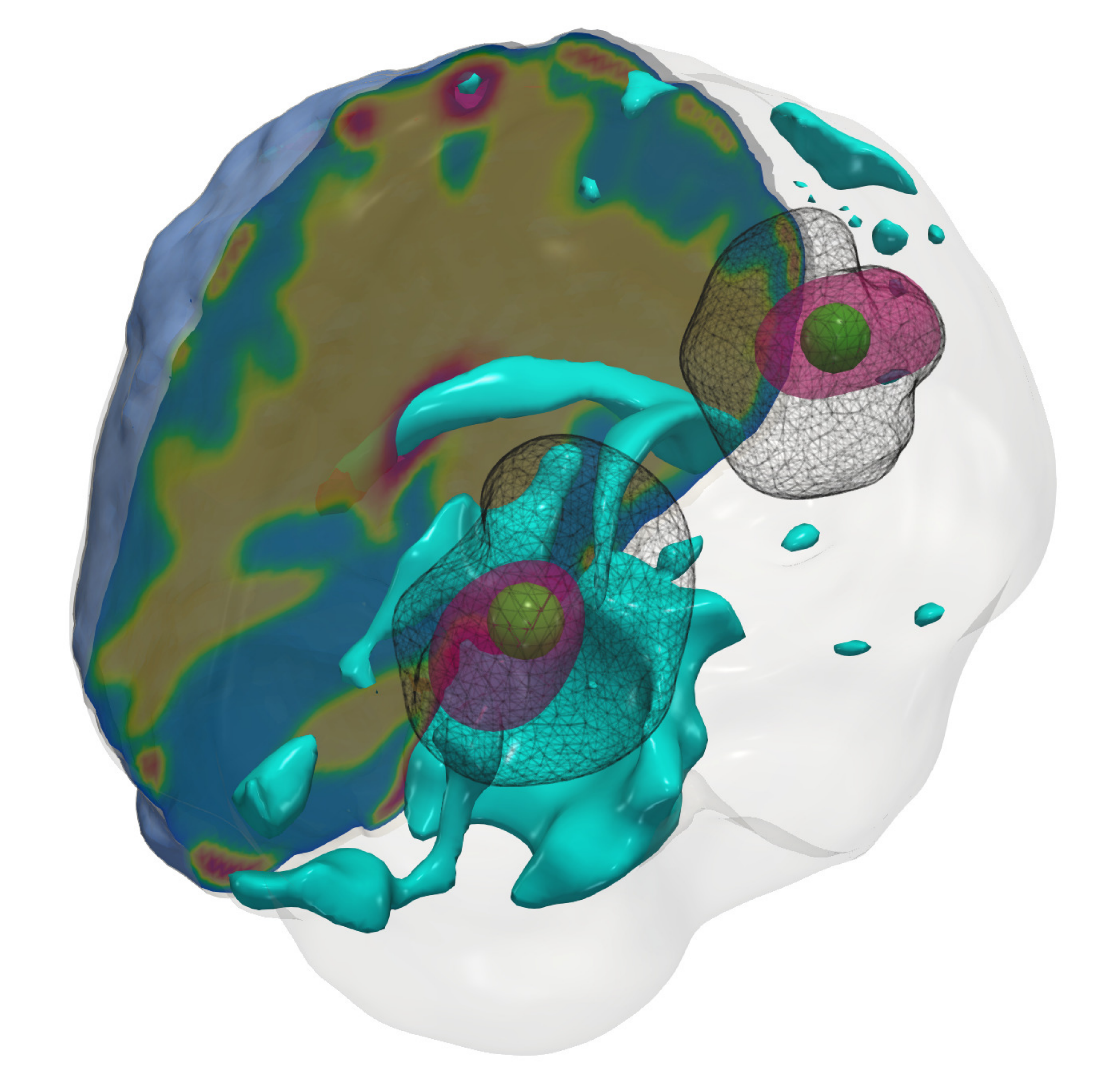}}%
	\subfloat[\textbf{\textit{CS}}]{\includegraphics[width=0.45\textwidth]{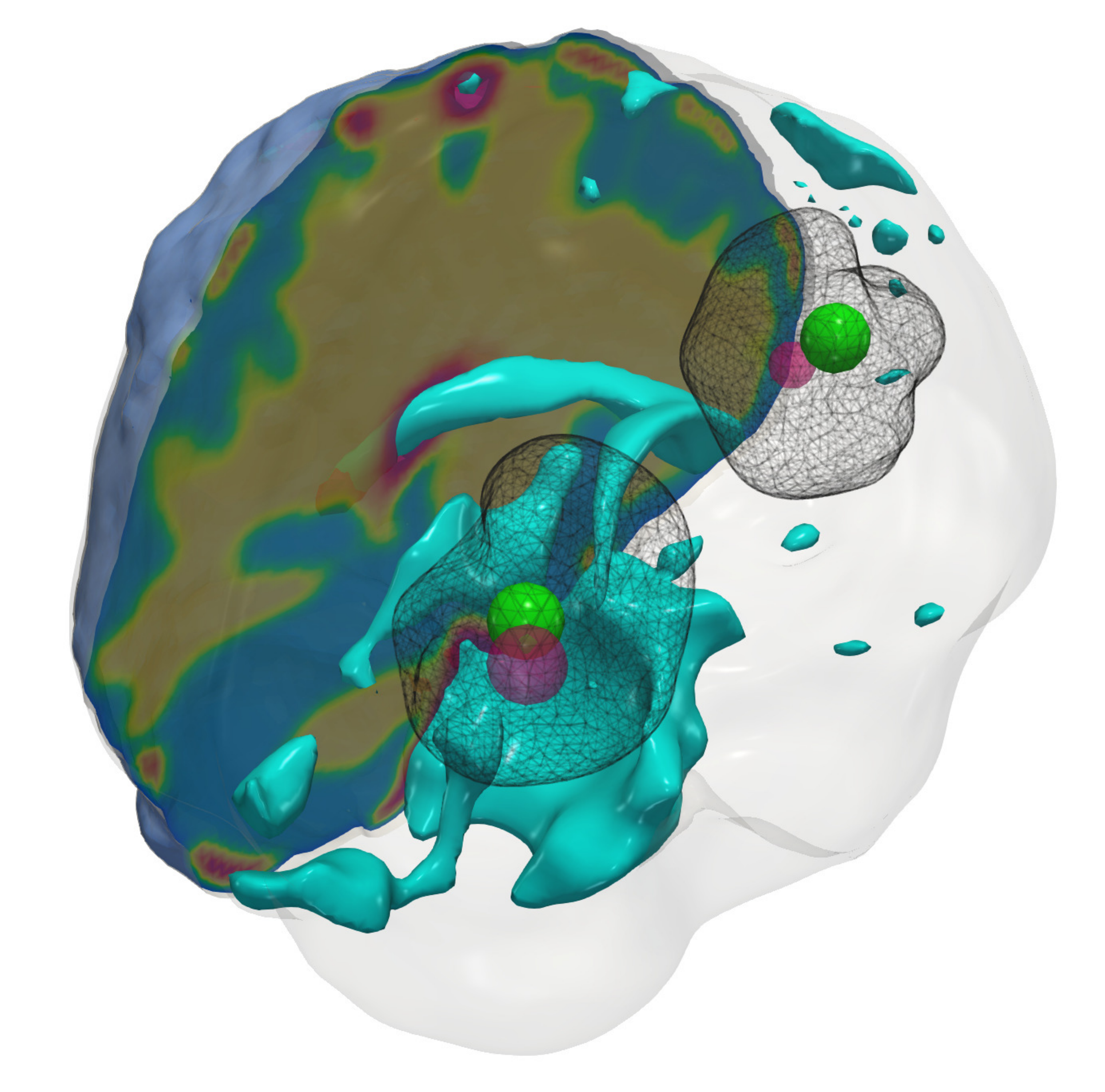}}
	\caption{\textit{Qualitative results for the \textbf{artificial tumor} test-case \textbf{\textit{AT-C4}} (multi-focal tumor) with growth parameters $\rho^\star = 8,~\kappa^\star = 0.025$.  The images show the tumor data (gray wireframe), ground truth tumor initial condition (green volume), reconstructed tumor initial condition (magenta volume), ventricles (cyan volume), and a section of the patient brain geometry. Both solvers are able to identify multi-focal initial conditions with the \textbf{\textit{CS}} solver enforcing more sparsity and hence obtaining better biophysical inverted parameters. As before, due to the ill-conditioning of the inverse problem, the exact locations of tumor initiation are not recovered.}\label{fig:at_c4}}
\end{figure}

\subsection{Artificial tumor (AT) test-case}\label{sec:AT}
In this test-case, the data $d(\vect{x})$ is generated using the non-linear forward model with inhomogeneous coefficients to capture the heterogeneity of tumors. These experiments are intended as a proof-of-concept of the performance of our solver on 3D synthetic datasets.

\medskip \noindent \textit{Setup. }We grow synthetic tumors in the segmentation of a statistical brain atlas~\cite{cocosco1997brainweb} MRI scan (an atlas obtain by averaging several MR images). The segmentation labels are gray matter, white matter, and cerebrospinal fluid-filled ventricles (CSF). We use sparse initial conditions to grow the tumor to a significant size resembling clinical observations and invert for all biophysical parameters using the grown tumor at $t = 1$ as input data to the solver. For all our test cases, we assume that the tumor grows and diffuses only in white matter. For this test-case, we observe the data everywhere, i.e., $c_d = 0$. We consider the following variations: \\

\begin{tabular}{lllll}
\textit{(i)}    & \textit{AT-C1: } &  medium sized, mono-focal tumor   & $\rho^{\star}=8$  & $\kappa^{\star}=0.025$   \\
\textit{(ii)}   & \textit{AT-C2: } &  large sized,  mono-focal tumor   & $\rho^{\star}=12$ & $\kappa^{\star}=0.05$    \\
\textit{(iii)}  & \textit{AT-C3: } &  multi-focal tumor, nearby seeds  & $\rho^{\star}=10$ & $\kappa^{\star}=0.025$   \\
\textit{(iv)}   & \textit{AT-C4: } &  multi-focal tumor, distant seeds & $\rho^{\star}=8$  & $\kappa^{\star}=0.025$   \\
\end{tabular} \\

\noindent We report our quantitative results (performance measures, convergence, and solver timings) in~\tabref{tab:synthetic_tc} and visualize the reconstruction to qualitatively assess the performance of our solvers in~\figref{fig:at_c1} -~\figref{fig:at_c4}. Each figure shows an isometric view of the patient brain with 3D volumes of the grown tumor data, ground truth, and reconstructed initial condition. We additionally visualize the ventricles and sections of healthy tissue to reveal the heterogeneity of the brain (and hence the tumor).

\medskip \noindent \textit{Observations. }We observe that for the medium sized tumor (\textit{AT-C1}) the reconstruction is nearly perfect for the \textbf{\textit{CS}} solver while the \textit{$\mathbold{L_2}$} solver produces a lower performance, particularly in the inversion of the reaction coefficient. This is primarily due to the rich initial conditions produced by the \textit{$\mathbold{L_2}$} solver which impedes its ability to predict the correct reaction scaling using our method. We note that if the reaction scaling is known beforehand, the \textit{$\mathbold{L_2}$} solver can potentially have better performance (see~\cite{Scheufele:2017a, Gholami:2017a} for similar synthetic experiments). This problem is magnified for larger tumors (\textit{AT-C2} and \textit{AT-C3}) where the predicted reaction coefficient shows about $66\%$ and $54\%$ relative error in the \textit{$\mathbold{L_2}$} solver for the two test-cases respectively (as compared to around $15\%$ and $1\%$ error with sparsity constraints). The reconstruction in diffusivity also suffers similarly without enforcing sparsity. The final test-case shows a multi-focal tumor with two far apart proliferation sites. While both the solvers produce disjoint initial conditions for the tumor, the \textbf{\textit{CS}} solver does better in reconstructing the model parameters: 2\% versus 37\% relative error in diffusivity and 9\% versus 50\% relative error in reaction coefficient. Both solvers show similar performance in their ability to fit the data (as measured by the relative error in final tumor reconstruction). Hence, our synthetic observations show consistent improvement in target parameter estimation for all test-cases with sparse initial conditions, without sacrificing the final tumor reconstruction quality.
{
	\setlength\tabcolsep{6pt}
	\begin{table}
		\caption{\textit{
				Quantitative results for the \textbf{artificial tumor} test-case in which we compare the \textit{$\mathbold{L_2}$} solver and the \textbf{\textit{CS}} solver. The ground truth forward solver parameters are indicated by '$^\star$' and their respective reconstructed values by '$\text{rec}$'. We also report the relative error in the inverted coefficient values ($e_{\kappa}, e_{\rho}$), the relative error in the final tumor reconstruction ($\elltwoT$), the relative error in the reconstructed initial condition ($\elltwoIC$), the relative error in the parameterization of the initial condition ($\elltwoICP$), the relative change in the gradient ($\relG$) and solver timings ($T_{inv}$) in seconds.
                 All simulations were run using 64 MPI tasks on 3 nodes of the Skylake partition of the Stampede2 system at TACC.}
			\label{tab:synthetic_tc}
		}
		\centering
		\begin{scriptsize}
			\makebox[\textwidth]{\centering
				\begin{tabular}{cccc|cLcLcLLcc}
					\toprule
					Test case & Solver & $\kappa^\star$ & $\rho^\star$ & $\kappa^\text{rec}$ &  $e_{\kappa}$ & $\rho^\text{rec}$ & $e_{\rho}$ & $\elltwoT$ & $\elltwoIC$ & $\elltwoICP$ & $\relG$ & $T_{inv}$\,[s]  \\
					\midrule

					\multirow{2}{*}{{\scriptsize \centering\textbf{\textit{AT-C1}}}}  &  \textit{$\mathbold{L_2}$} &  \num{2.5E-2} &	\num{8E0} &  \num{2.32071E-2}   & \num{7.1716E-2}	 & \num{4.07909E0}	& \num{4.901E-1}	 & 	\num{3.1304E-1}& \num{2.91907E0} & \num{9.13428E0}		&	\num{4.023723386941e-03}    &  \num[scientific-notation=true]{231}	\\

					&  \textbf{\textit{CS}}  &  \num{2.5E-2} & \num{8E0} & \num{2.49762E-2}   & \num{9.5200E-04}	& \num{8.00559E0}	& \num{6.9875e-04}	 & 	\num{2.70706E-4}& \num{1.24406E-4} & \num[scientific-notation=true]{0.0250494}  & \num{1.745881825607e-07}	& \num[scientific-notation=true]{842}	\\

					\midrule

					\multirow{2}{*}{{\scriptsize \centering\textbf{\textit{AT-C2}}}}  &  \textit{$\mathbold{L_2}$} &  \num{5E-2}  &	\num{1.2E1}  & \num{2.13462E-2}   & \num{5.731E-1}	& \num{4.13435E0}	& \num{6.555E-1}	 & 	\num{1.58267E-1} & \num{5.65916E0}	& \num{2.34728E1}		&  \num{7.388352546627e-03}	& \num[scientific-notation=true]{238}	\\

					&  \textbf{\textit{CS}}  &  \num{5E-2} &	\num{1.2E1}  & \num{3.88688E-2}   & \num{2.226E-01}	& \num{1.02395E1}	& \num{1.467E-01}	 & 	\num{1.7187E-1}& \num{1.26156E0}  & \num{2.25297E0}     &	 \num{9.364987434304e-06} & \num[scientific-notation=true]{643}	\\

					\midrule

					\multirow{2}{*}{{\scriptsize \centering\textbf{\textit{AT-C3}}}}  &  \textit{$\mathbold{L_2}$} &  \num{2.5E-2}  &	\num{1E1}  & \num{1.35671E-2}   & \num{4.573E-1}	& \num{4.61565E0}	& \num{5.384E-1}	 & 	\num{1.84325E-1} & \num{3.54952E0}	& \num{1.09151E1}  & \num{3.164418585704e-04}	& \num[scientific-notation=true]{321}	\\

					&  \textbf{\textit{CS}}  &  \num{2.5E-2} &	\num{1E1} & \num{2.67104E-2}   & \num{6.840E-02}	 & \num{9.9136E0}	& \num{8.6E-03}	 & 	\num{1.40215E-1}& \num{6.91656E-1} & \num{5.34797E-1}   & \num{2.081215823352e-06} &	\num[scientific-notation=true]{947}	\\

					\midrule

					\multirow{2}{*}{{\scriptsize \centering\textbf{\textit{AT-C4}}}}  &  \textit{$\mathbold{L_2}$} &  \num{2.5E-2} &	\num{8E0} & \num{1.56874E-2}   & \num{3.725E-1}	 & \num{3.95276E0}	& \num{5.059E-1}	 & 	\num{2.33584E-1} & \num{2.42075E0}	& \num{8.23165E0}  & \num{2.637048689932e-03} & 	\num[scientific-notation=true]{231}	\\

					&  \textbf{\textit{CS}}  &  \num{2.5E-2} &	\num{8E0} & \num{2.56612E-2}   & \num{2.64E-02} & \num{7.30685E0}	& \num{8.66E-2}	 & 	\num{2.91632E-1}& \num{1.12572E0} & \num{2.00598E0} &	\num{5.307858377507e-06 } & \num[scientific-notation=true]{1407}	\\

					\bottomrule
			\end{tabular}}
		\end{scriptsize}
	\end{table}
}

\subsection{Artificial tumor with observation operator test-case}
As mentioned before, an observation operator is necessary since the full extent of tumor invasion is not detectable using existing imaging technologies. The purpose of this experiment is to investigate the effect of the observation operator $\mat{O}$ on the model parameter reconstruction abilities of the solver. In particular, we would like to see if partial observations significantly deteriorate the reconstruction quality.

\medskip \noindent \textit{Setup. }We assess the performance and sensitivity of our solver by choosing different values for the detection threshold $c_d \in \{0, 0.1, 0.2, 0.3, 0.4, 0.5\}$, which defines the observation points according to Eq.~\eqref{e:obs}.  We define the parameter $n_d$ to quantify the number of voxels observed with each threshold value. We choose the synthetic test-case $\textit{AT-C2}$ as a suitable representation of a large and significantly invasive (aggressive) tumor (large parameter values: $\rho^\star = 12$, $\kappa^\star = \num{5E-2}$).  We report all our performance metrics for the different observation thresholds in~\tabref{tab:obs}. We emphasize that the final tumor reconstruction error is with respect to the fully grown tumor with the ground truth model parameters and not the data restricted to the observation points. This definition is more useful in helping us understand if the model is reconstructing the full extent of the tumor, including the areas which are typically invisible in imaging data.

\medskip \noindent \textit{Observations. }The most important observation is that the solver is able to approximately establish the extent of tumor invasion up to a detection threshold of $c_d = 0.3$. This is reflected in the  performance metrics that do not register significant drops over these threshold values. The final tumor reconstruction error shows perturbations of up to 4\% till this detection threshold. The values for the inverted model parameters also do not exhibit appreciable degradation and roughly remain around the ground truth coefficients -- $\num{9.8}$ to $\num{11.1}$ for the reaction coefficient and $\num[scientific-notation=true]{0.0389}$ to $\num[scientific-notation=true]{0.06}$ for the diffusion coefficient. Larger detection thresholds (from $c_d = 0.4$ onwards) lead to worsening reconstruction errors and significant deterioration in target parameter estimation. This is expected since there is insufficient data for any reliable inversion, especially since this tumor has substantial diffusion resulting in considerable loss of data when choosing higher values for the detection threshold.  Finally, while there is no consensus regarding the value of $c_d$ in literature (generally values around 0.2 - 0.4 have been used~\cite{konukoglu2010image, Swanson2008}), this value can easily be modified in our formulation once more specific information becomes available.

{
	\setlength\tabcolsep{6pt}
	\begin{table}
		\caption{\textit{
				Quantitative results for \textbf{different observation thresholds}, $c_d \in \{0, 0.1, 0.2, 0.3, 0.4, 0.5\}$. We choose the ground truth model parameters from the synthetic test-case \textbf{AT-C2} since they produce a large tumor. We report the performance metrics for model parameter inversion ($e_{\kappa}, e_{\rho}$), the relative error in the final tumor reconstruction ($\elltwoT$),  the relative error in the reconstructed initial condition ($\elltwoIC$), the relative error in the parameterization of the initial condition ($\elltwoICP$), the relative change in the gradient ($\relG$) and solver timings ($T_{inv}$) in seconds. All ground truth values are indicated by '$^\star$' and the inverted quantities with '$\text{rec}$'. We note that for $c_d = 0$ all voxels are observed and the results are replicated from~\tabref{tab:synthetic_tc}. The number of  voxels observed is denoted by $n_d$.
				\label{tab:obs}
		}}
		\centering
		\begin{scriptsize}
			\makebox[\textwidth]{\centering
				\begin{tabular}{cccc|cLcLcLLcc}
					\toprule
					$c_d$  & $n_d$ & $\kappa^\star$  & $\rho^\star$ & $\kappa^\text{rec}$ &  $e_{\kappa}$   & $\rho^\text{rec}$ & $e_{\rho}$ & $\elltwoT$ & $\elltwoIC$  & $\elltwoICP$ & $\relG$ & $T_{inv}$\,[s]  \\
					\midrule
					0 & - & \num{5E-2} &	\num{1.2E1}  & \num{3.88688E-2}   & \num{2.226E-01}	& \num{1.02395E1}	& \num{1.467E-01}	 & 	\num{1.7187E-1}& \num{1.26156E0}  & \num{2.25297E0}     &	 \num{9.364987434304e-06} & \num[scientific-notation=true]{643}	\\

                    \num{0.1} & \num[scientific-notation=true]{25434} & \num{5E-2} &	\num{1.2E1}  & \num{5.03163E-2}   & \num{6.3E-03}	& \num{9.80016E0}	& \num{1.833E-01}	 & 	\num{1.56335E-1}& \num{1.20199E0} &\num{2.10042E0}   &	 \num{1.004327037073e-06} & \num[scientific-notation=true]{1167}	\\

					\num{0.2} & \num[scientific-notation=true]{19994} & \num{5E-2} &	\num{1.2E1}  & \num{4.34648E-2}   & \num{1.37E-02}	& \num{11.1047E0}	& \num{7.46E-02}	 & 	\num{1.93852E-1}& \num{1.20156E0} &\num{2.05124E0}   &	 \num{9.714040852552e-06} & \num[scientific-notation=true]{771}	\\

                    \num{0.3} & \num[scientific-notation=true]{16692} & \num{5E-2} &	\num{1.2E1}  & \num{6.00351E-2}   & \num{2.007E-01}	& \num{9.95096E0}	& \num{1.708E-01}	 & 	\num{1.98078E-1}& \num{1.20161E0} &\num{2.09298E0}   &	 \num{6.871469860592e-06} & \num[scientific-notation=true]{1165}	\\

					\num{0.4} &  \num[scientific-notation=true]{14135} & \num{5E-2} &	\num{1.2E1}  & \num{7.11374E-2}   & \num{4.227E-1}	& \num{9.90314E0}	& \num{1.747E-1}	 & 	\num{3.19771E-1}& \num{1.26636E0}   &	\num{3.42173E0} & \num{3.173672512967e-06} & \num[scientific-notation=true]{1181}	\\

                    \num{0.5} &  \num[scientific-notation=true]{11972} & \num{5E-2} &	\num{1.2E1}  & \num{8.16838E-2}   & \num{6.337E-1}	& \num{8.83878E0}	& \num{2.634E-1}	 & 	\num{3.16673E-1}& \num{1.32881E0}   &	\num{2.7034E0} & \num{2.093170834297e-06} & \num[scientific-notation=true]{959}	\\
					\bottomrule
			\end{tabular}}
		\end{scriptsize}
	\end{table}
}

\subsection{Artificial tumor with varying spatial resolution test-case}
The ground truth for parameter estimation should ideally be independent of the numerical discretization used. Since we use a pseudo-spectral discretization scheme which demands smooth functions (to avoid aliasing errors), a mesh-dependent numerical diffusion is introduced in our formulation. The purpose of this test-case is to assess the degree of information loss due to spatial discretization and to understand the extent of reliability of our inverted parameters with mesh size.

\medskip \noindent \textit{Setup: }We consider the synthetic test-case \textit{AT-C2} as a suitable representation of a large and invasive tumor. Since there is no analytical solution to the forward problem, we grow a synthetic tumor in a high spatial resolution of $512^3$ and assume this to be the ``true" tumor. We sub-sample this tumor to resolutions $128^3$ and $256^3$ and use the resulting tumors as data for inversion in the respective spatial discretizations. The initial radial basis $\sigma$ is kept constant across all grids (set to the spacing of the $128^3$ grid). We consider the following test cases: \\

\begin{tabular}{llll}
\textit{(i)}   & \textit{C1: } &  $\rho^{\star}=12$  & $\kappa^{\star}=0$   \\
\textit{(ii)}  & \textit{C2: } &  $\rho^{\star}=12$  & $\kappa^{\star}=\num{5E-4}$    \\
\textit{(iii)} & \textit{C3: } &  $\rho^{\star}=12$  & $\kappa^{\star}=\num{5E-3}$   \\
\textit{(iv)}  & \textit{C4: } &  $\rho^{\star}=12$  & $\kappa^{\star}=\num{5E-2}$   \\
\end{tabular}\\

We use different orders of magnitude for the diffusion coefficient to gauge the extent of numerical diffusion, relative to actual physical diffusion. We report our performance measures in~\tabref{tab:spatial}.

\medskip \noindent \textit{Observations: }We observe that we cannot recover diffusion coefficients smaller than around $\num{1E-3}$ using a spatial resolution of $128^3$. This can be seen in the values for $\kappa^\text{rec}$ (diffusivity reconstruction) which fail to recover any smaller values (test-case \textit{C1}; \textit{C2}). As a result, the values for $\rho^\text{rec}$ (reconstructed reaction magnitude) also improve for test-cases with higher ground truth diffusivity values as the effect of mesh-dependent diffusion is reduced. The higher resolution mesh ($256^3$) is able to recover diffusion coefficient values up to an order of magnitude of $\num{1E-5}$.
Consequently, we observe larger differences for the $\rho^\text{rec}$ values between the two resolutions as the ground truth diffusivity drops. The higher resolution inversion does a better job in estimating the diffusivity (due to lower mesh-diffusion errors) and hence can estimate the reaction coefficient reasonably where the lower resolution inversion performs poorly.
Hence, our estimated parameters are well founded only if they are not in the range of mesh-induced numerical diffusion. One way to circumvent this error is to perform grid continuation where we invert for parameters in successively higher mesh resolutions using the sparsity and inverted parameters from the coarser meshes as an initial guess and settle on a resolution as the results of the inversion stagnate.

{
    \setlength\tabcolsep{6pt}
    \begin{table}
        \caption{\textit{
                Quantitative results for \textbf{varying spatial resolution}. We report the performance metrics for model parameter inversion ($e_{\kappa}, e_{\rho}$), the relative error in the final tumor reconstruction ($\elltwoT$),  the relative error in the reconstructed initial condition ($\elltwoIC$), the relative error in the parameterization of the initial condition ($\elltwoICP$), the relative change in the gradient ($\relG$). All ground truth values are indicated by '$^\star$' and the inverted quantities with '$\text{rec}$'. For the higher resolution ($256^3$), we use 128 MPI tasks on 6 nodes. We do not report the relative error for inverted diffusivity in test-case \textbf{C1} since the ground truth is zero}.}\label{tab:spatial}
        \centering
        \begin{scriptsize}
            \makebox[\textwidth]{\centering
                \begin{tabular}{cccc|LcLccLLc}
                    \toprule
                    Test-case & Spatial resolution  & $\kappa^\star$  & $\rho^\star$ & $\kappa^\text{rec}$ &  $e_{\kappa}$   & $\rho^\text{rec}$ & $e_{\rho}$ & $\elltwoT$ & $\elltwoIC$  & $\elltwoICP$ & $\relG$  \\
                    \midrule
                    \multirow{2}{*}{{\scriptsize \centering\textbf{\textit{C1}}}} &   $256^3$ &  \num{0} &	\num{1.2E1}  & \num{9.21335e-06}   & -	& \num{1.07212E1}	& \num{1.066E-01}	 & 	\num{1.53553E-1}& \num{1.49372E-1}   &	 \num{8.39195E-2} & \num{1.681966204404e-06}	\\
                    & $128^3$ &  \num{0} &	\num{1.2E1}  & \num{1.87839E-3}   & -	& \num{7.28431E0}	& \num{3.93E-01}	 & 	\num{3.60323E-1}& \num{1.30274E0}   &	 \num{1.64031E0} & \num{3.810642746152e-06}	\\
                    \midrule
                    \multirow{2}{*}{{\scriptsize \centering\textbf{\textit{C2}}}} &   $256^3$ &  \num{5e-4} &	\num{1.2E1}  & \num{7.7636E-4}  & \num{5.527E-1}	& \num{9.77093E0}	& \num{1.858E-01}	 & 	\num{1.67888E-1}& \num{4.10647E-1}   &	 \num{5.00565E-1} & \num{4.636558128288e-06}	\\
                    & $128^3$ &  \num{5e-4} &	\num{1.2E1}  & \num{3.67746E-3}   & \num{6.3549E0}	& \num{7.19341E0}	& \num{4.005E-01}	 & 	\num{3.85076E-1}& \num{1.77019E0}   &	 \num{2.46416E0} & \num{1.852486670687e-06}	\\
                    \midrule
                    \multirow{2}{*}{{\scriptsize \centering\textbf{\textit{C3}}}} &   $256^3$ &  \num{5E-3} &	\num{1.2E1}  & \num{3.53552E-3}   & \num{2.929E-1}	& \num{8.60731E0}	& \num{2.827E-01}	 & 	\num{1.42012E-1}& \num{1.47645E0}   &	 \num{3.2121E0} & \num{4.045040605829e-06}	\\
                    & $128^3$ &  \num{5E-3} &	\num{1.2E1}  & \num{3.84162E-3}   & \num{2.317E-1}	& \num{8.52066E0}	& \num{2.899E-01}	 & 	\num{2.22659E-1}& \num{1.34372E0}   &	 \num{1.94505E0} & \num{8.296471557819e-06}	\\
                    \midrule
                    \multirow{2}{*}{{\scriptsize \centering\textbf{\textit{C4}}}} &   $256^3$ &  \num{5E-2} &	\num{1.2E1}  & \num{4.36627E-2}   & \num{1.267E-1}	& \num{9.80192E0}	& \num{1.832E-01}	 & 	\num{1.79256E-1}& \num{1.45005E0}   &	 \num{2.52883E0} & \num{7.505800785923e-06}	\\
                    & $128^3$ &  \num{5E-2} &	\num{1.2E1}  & \num{3.02396E-2}   & \num{3.952E-1}	& \num{1.08961E1}	& \num{9.2E-02}	 & 	\num{2.6926E-1}& \num{1.6043E0}   &	 \num{2.2124E0} & \num{4.957205997243e-06}	\\
                    \bottomrule
            \end{tabular}}
        \end{scriptsize}
    \end{table}
}

\subsection{Artificial tumor with noise test-case}
The purpose of this test-case is to examine the robustness of our inversion against noise in the data. Instead of using high frequency uncorrelated noise (white noise), we are interested in observing the effects of a more realistic low-frequency noise model.

\medskip \noindent \textit{Setup. }We approximate the source of noise by modifying the low-frequency spectrum of our data. We add a uniform random noise scaled by the inverse squared frequency to the respective frequency components of the data . Then, we take the inverse Fourier transform of this noisy spectrum to obtain our noise-corrupted data. We use AccFFT to compute our frequency spectrum and IFFTs in a distributed manner. We choose our ground truth parameters from the artificial tumor test-case \textit{AT-C2} since this test-case features the largest invasive synthetic tumor making the inversion most ill-conditioned. We add various levels of noise (as measured by the relative error in the true data and noise-corrupted data) and report our findings in~\tabref{tab:noise}.

\medskip \noindent \textit{Observations. }We observe that our model does not exhibit any significant sensitivity for this noise model. We consider the small changes in performance and convergence metrics between the different noise levels inconsequential and attribute them to the inherent randomness in generating the data.

{
	\setlength\tabcolsep{6pt}
	\begin{table}
		\caption{\textit{
				Quantitative results for \textbf{different low frequency noise levels}. We choose the ground truth model parameters from the synthetic test-case \textbf{AT-C2}. We report the performance metrics for model parameter inversion ($e_{\kappa}, e_{\rho}$), the relative error in the final tumor reconstruction ($\elltwoT$),  the relative error in the reconstructed initial condition ($\elltwoIC$), the relative error in the parameterization of the initial condition ($\elltwoICP$), the relative change in the gradient ($\relG$) and solver timings ($T_{inv}$) in seconds. All ground truth values are indicated by '$^\star$' and the inverted quantities with '$\text{rec}$'.
				\label{tab:noise}
		}}
		\centering
		\begin{scriptsize}
			\makebox[\textwidth]{\centering
				\begin{tabular}{ccc|cLcLcLLcc}
					\toprule
					Noise level  & $\kappa^\star$  & $\rho^\star$ & $\kappa^\text{rec}$ &  $e_{\kappa}$   & $\rho^\text{rec}$ & $e_{\rho}$ & $\elltwoT$ & $\elltwoIC$  & $\elltwoICP$ & $\relG$ & $T_{inv}$\,[s]  \\
					\midrule
					0\% &  \num{5E-2} &	\num{1.2E1}  & \num{3.88688E-2}   & \num{2.226E-01}	& \num{1.02395E1}	& \num{1.467E-01}	 & 	\num{1.7187E-1}& \num{1.26156E0}  & \num{2.25297E0}     &	 \num{9.364987434304e-06} & \num[scientific-notation=true]{643}	\\
					1\% &  \num{5E-2} &	\num{1.2E1}  & \num{3.7047E-2}   & \num{2.591E-1}	& \num{1.02919E1}	& \num{1.423E-1}	 & 	\num{1.86536E-1}& \num{1.20203E0}   &	\num{2.09698E0} & \num{3.559274745038e-03} & \num[scientific-notation=true]{996}	\\
					5\% &  \num{5E-2} &	\num{1.2E1}  & \num{3.73551E-2}   & \num{2.529E-1}	& \num{1.0923E1}	& \num{8.97E-2}	 & 	\num{1.72374E-1}& \num{1.20191E0} & \num{2.09139E0}   &	 \num{7.915077693785e-06} & \num[scientific-notation=true]{788}	\\
					10\% &  \num{5E-2} &	\num{1.2E1}  & \num{4.07252E-2}   & \num{1.855E-1}	& \num{1.08886E1}	& \num{9.26E-2}	 & 	\num{1.72522E-1}& \num{1.20167E0}   & \num{2.05966E0}	& \num{3.107339180313e-04} & \num[scientific-notation=true]{1088}	\\
					\bottomrule
			\end{tabular}}
		\end{scriptsize}
	\end{table}
}

\subsection{Real tumor (RT) test-case}\label{sec:rt}
The purpose of this test-case is to evaluate our solver on real patient brains diagnosed with glioblastoma. While we have no knowledge of the ground truth model parameters for real data, this experiment attempts to demonstrate the similarity in reconstruction trends observed in the artificial tumor test-cases (as designed above) and a clinical dataset.

\begin{figure}
	\subfloat[\textit{$\mathbold{L_2}$ }]{\includegraphics[width=0.45\textwidth]{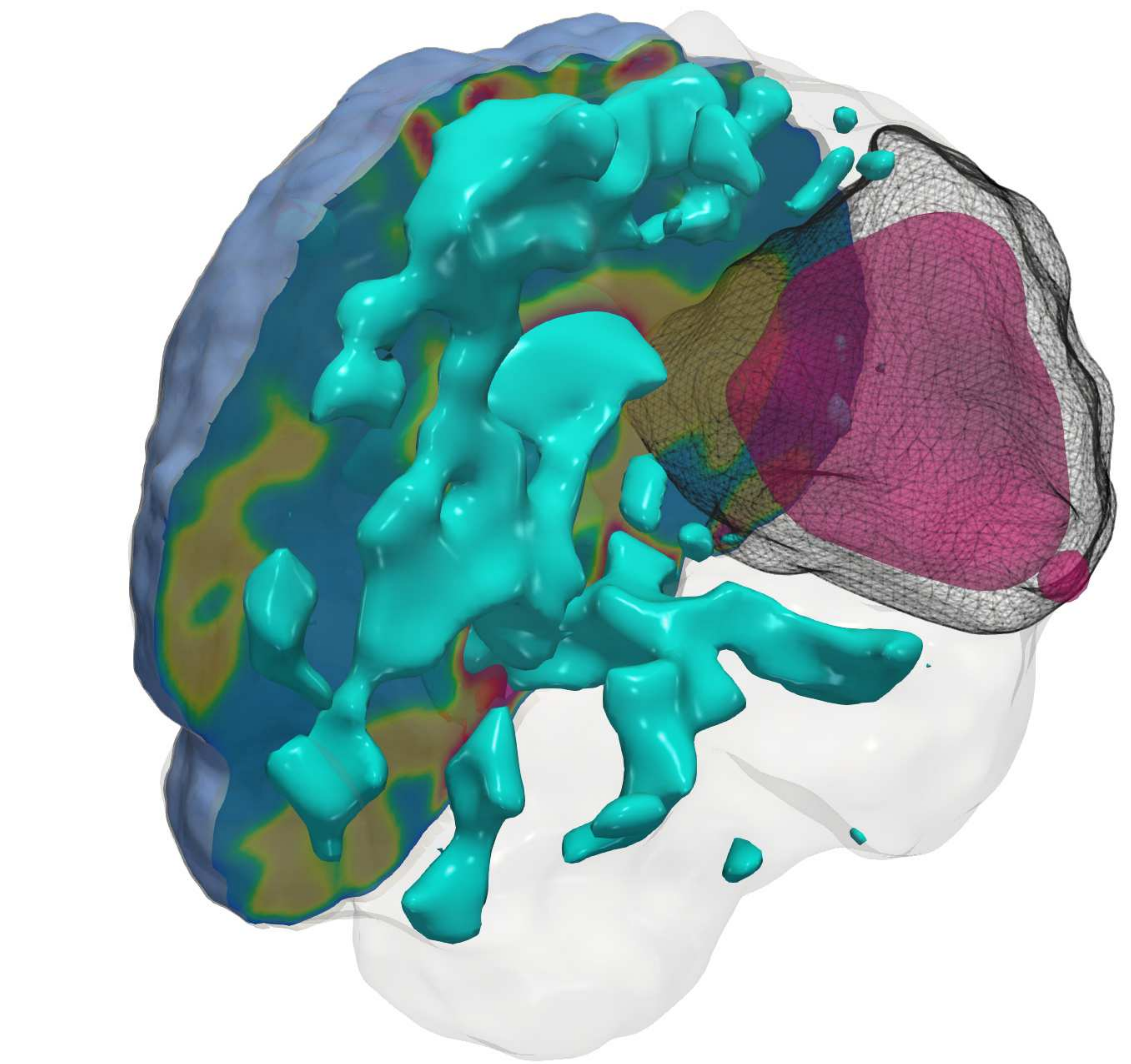}}%
	\subfloat[\textbf{\textit{CS}}]{\includegraphics[width=0.45\textwidth]{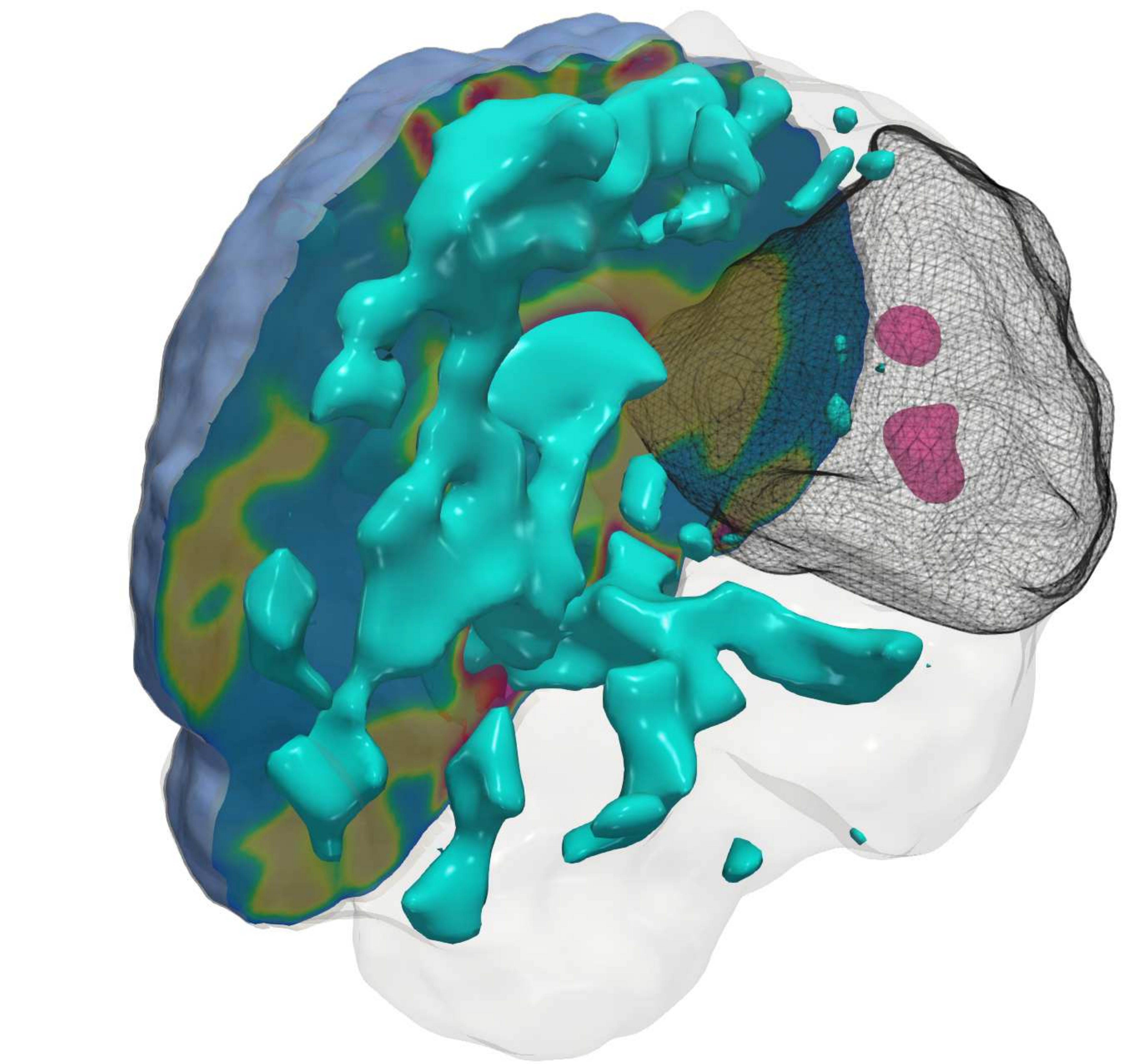}}
	\caption{\textit{Qualitative results for a \textbf{real tumor} taken from the BraTS'18 TCIA dataset. The images show the tumor core (enhancing and necrotic tumor cells) indicated as a gray wireframe with the reconstructed initial condition (magenta volume) and parts of the patient brain geometery. We observe similar trends as the synthetic tumor test-cases with sparse tumor initial conditions for the \textbf{\textit{CS}} solver which predicts higher diffusion and reaction coefficients to fit the tumor data.}\label{fig:real}}
\end{figure}

\medskip \noindent \textit{Setup: }Since we do not have access to the healthy patient brain, we cannot resort to direct inversion of model parameters. There have been many approaches in literature (see~\cite{Gholami:2017a, Scheufele:2017a} for a comprehensive review) to approximate the healthy brain through methods such as diffeomorphic image registration and registration-tumor coupled optimization formulations. While such techniques are necessary to capture the healthy patient brain, we do not pursue them since this test-case is intended to primarily serve as an illustration. Instead, we simply replace all tumor segmented voxels in the patient image with white matter and invert only for the diffusivity and reaction in the white matter. We note that by doing so we have lost any information concerning the inhomogeneity of our model coefficients in white and gray matter. We use MRI scans from the BraTS dataset~\cite{Menze:2015a}. These images are annotated with voxel-wise segmentation of tumor acquired through manual delineation performed by expert radiologists. For the healthy tissue segmentation, we refer to~\cite{brats18} where the brain is further segmented into gray matter, white matter and cerebrospinal fluid-filled ventricles through diffeomorphic image registration with healthy atlases. Since our model incorporates only a single species of tumor cells, we combine the enhancing and necrotic tumor sub-classes into one tumor class  (also known as the tumor core) and use this as our tumorous species. We report the inverted values for our model coefficients, data reconstruction error, and convergence metrics in~\tabref{tab:real_tc}. We visualize our reconstruction in~\figref{fig:real}.

\medskip \noindent \textit{Observations: }We observe sparse initial condition reconstruction with the \textbf{\textit{CS}} solver and consequently larger diffusion and reaction coefficients for the tumor model. This is in accordance with our observations in the synthetic tumor test-cases. Both solvers produce roughly equivalent reconstruction errors for the final tumor condition which is around 16-19\%. We note that if our primary goal is to reduce this mismatch we might fare better with more complex initial conditions which resemble fully grown tumors. At the extreme scenario, pure interpolation of the data with radial basis, i.e., setting all model coefficients to zero, would lead to a significantly better mismatch. However, this incurs the severe penalty of losing all information about the biophysical model at hand. Since the focus of this work is on the latter, we accept higher reconstruction errors as long as we can identify the target model parameters reliably. This is ultimately more useful for tumor growth understanding and prediction. We emphasize that approaches like a coupled image registration and tumor inversion formulation are necessary for access to a more informative healthy patient (see~\cite{Scheufele:2017a} for more details). Our plan is to conduct a systematic study of the solver considering hundreds of cases. However, such a study is beyond the scope of this paper and requires a separate treatment.

{
	\setlength\tabcolsep{6pt}
	\begin{table}
		\caption{ \textit{
			Quantitative results for a \textbf{real patient image} taken from the BraTS dataset~\cite{menze2015multimodal}. Since it is impossible to know the ground truth parameter values, we only report the inverted model coefficient values indicated by '$\text{rec}$' and the final tumor reconstruction error with respect to the data ($\elltwoT$). We additionally report the relative gradient change ($\relG$) and solver timings ($T_{{inv}}$) in seconds. }
			\label{tab:real_tc}
		}
		\centering
		\begin{scriptsize}
			\makebox[\textwidth]{\centering
				\begin{tabular}{cc|LLLc|c}
					\toprule
					Brain ID & Solver   & $\kappa^\text{rec}$  & $\rho^\text{rec}$  & $\elltwoT$  &  $\relG$  & $T_{inv}$\,[s]  \\
					\midrule

					\multirow{2}{*}{{\scriptsize \centering\textbf{\textit{BRATS-TCIA}}}}  &  \textit{$\mathbold{L_2}$} & \num{6.99004E-3}  & \num{4.36271E0} & \num{1.56418E-1} & \num{2.894303416166e-04} & \num{3.20E2}  	\\

					&  \textbf{\textit{CS}} & \num{3.02798E-2}  & \num{9.76668E1} & \num{1.8595E-1} & \num{6.464630304957e-06} & \num{7.57E2}  	\\

					\bottomrule
			\end{tabular}}
		\end{scriptsize}
	\end{table}
}

\section{Conclusions}
\label{sec:conclusions}
We presented an inverse problem formulation for the simultaneous estimation of all parameters in a commonly used reaction-diffusion model, namely the reaction coefficient, diffusion coefficient, and the spatial distribution of the tumor initial condition. The data that drive the inversion come from MRI scans, in particular the segmentation of tumor regions. 
We motivated the need for the sparse localization of the initial tumor in order to reliably reconstruct the other parameters. Additionally, this constraint is biophysically motivated and hence can be of clinical relevance in understanding the origin of the tumor and its role in the subsequent evolution of the disease.  We presented results demonstrating the performance of our solver in reconstructing the ground truth model parameters, even for large and complex synthetic tumors. The results also highlight significant improvement in reconstruction quality with the use of sparsity constraints. We observe similar reconstruction trends in real patient data, thus providing a proof-of-concept of applicability in clinical practice.
Finally, we employed novel reduced-space optimization algorithms and compressive sampling to solve our non-linear inverse problem and have tested its robustness against partial data observations and noise. Our software is implemented in a scalable manner enabling realistic time-to-solutions, a significant factor for applications in a diagnostic setting. While limitations like the necessity of more complex growth models exist, we envision this work as an important stepping stone towards the personalization of tumor growth models.

Our next step is to conduct a  comprehensive evaluation of the model using a large number of clinical images. Our goal with such a study  will be to correlate the model parameters with clinical outcome, for example survival.  There are several ways that this work can be extended. Even if multiple time-snapshot scans exist, the presence of partial observations still poses difficulties. The same framework we discussed here can be used, but with multiple scans the problems of inverting for $\rho$ and $\kappa$ and the tumor initiation can be possibly decoupled.  A more significant challenge is incorporating more complex biophysical models, while still having a single scan. All these directions are ongoing work in our group.

\section*{Acknowledgments}
This material is based upon work supported by NIH award 5R01NS042645-14; by NSF awards CCF-1817048 and CCF-1725743;  by the U.S. Department of Energy, Office of Science, Office of Advanced Scientific Computing Research, Applied Mathematics program under Award Number DE-SC0019393; and by the U.S. Air Force Office of Scientific Research award FA9550-17-1-0190.  Any opinions, findings, and conclusions or recommendations expressed herein are those of the authors and do not necessarily reflect the views of the AFOSR,  DOE, NIH, and NSF. Computing time on the Texas Advanced Computing Centers Stampede system was provided by an allocation from TACC and the NSF.

%
%
%
%
\bibliography{literature}

\begin{thebibliography}{10}

\bibitem{Bakas17}
{\sc S.~Bakas, H.~Akbari, A.~Sotiras, M.~Bilello, M.~Rozycki, J.~S. Kirby,
  J.~B. Freymann, K.~Farahani, and C.~Davatzikos}, {\em Advancing the cancer
  genome atlas glioma mri collections with expert segmentation labels and
  radiomic features}, Scientific Data, 4 (2017).

\bibitem{Beck:2009}
{\sc A.~Beck and M.~Teboulle}, {\em A fast iterative shrinkage-thresholding
  algorithm for linear inverse problems}, SIAM J. Img. Sci., 2 (2009),
  pp.~183--202.

\bibitem{Blumensath:2009}
{\sc T.~Blumensath and M.~E. Davies}, {\em Iterative hard thresholding for
  compressed sensing}, Applied and Computational Harmonic Analysis, 27 (2009),
  pp.~265 -- 274.

\bibitem{Candes:2006}
{\sc E.~J. {Candes}, J.~{Romberg}, and T.~{Tao}}, {\em Robust uncertainty
  principles: exact signal reconstruction from highly incomplete frequency
  information}, IEEE Transactions on Information Theory, 52 (2006),
  pp.~489--509.

\bibitem{Castillo60}
{\sc M.~Castillo, J.~K. Smith, L.~Kwock, and K.~Wilber}, {\em Apparent
  diffusion coefficients in the evaluation of high-grade cerebral gliomas},
  American Journal of Neuroradiology, 22 (2001), pp.~60--64.

\bibitem{Chen:2012b}
{\sc X.~Chen, R.~M. Summers, and J.~Yoa}, {\em Kidney tumor growth prediction
  by coupling reaction-diffusion and biomechanical model}, Biomedical
  Engineering, IEEE Transactions on, 60 (2012), pp.~169--173.

\bibitem{clatz2005realistic}
{\sc O.~Clatz, M.~Sermesant, P.-Y. Bondiau, H.~Delingette, S.~K. Warfield,
  G.~Malandain, and N.~Ayache}, {\em Realistic simulation of the 3-d growth of
  brain tumors in {MR} images coupling diffusion with biomechanical
  deformation}, Medical Imaging, IEEE Transactions on, 24 (2005),
  pp.~1334--1346.

\bibitem{cocosco1997brainweb}
{\sc C.~A. Cocosco, V.~Kollokian, R.~K.-S. Kwan, G.~B. Pike, and A.~C. Evans},
  {\em Brainweb: Online interface to a 3d {MRI} simulated brain database}, in
  NeuroImage, Citeseer, 1997.

\bibitem{Colin:2014a}
{\sc T.~Colin, A.~Iollo, J.-B. Lagaert, and O.~Saut}, {\em An inverse problem
  for the recovery of the vascularization of a tumor}, Journal of Inverse and
  Ill-posed Problems, 22 (2014), pp.~759--786.

\bibitem{crank1996}
{\sc J.~Crank and P.~Nicolson}, {\em A practical method for numerical
  evaluation of solutions of partial differential equations of the
  heat-conduction type}, Advances in Computational Mathematics, 6 (1996),
  pp.~207--226.

\bibitem{Donoho:2012}
{\sc D.~L. {Donoho}, Y.~{Tsaig}, I.~{Drori}, and J.~{Starck}}, {\em Sparse
  solution of underdetermined systems of linear equations by stagewise
  orthogonal matching pursuit}, IEEE Transactions on Information Theory, 58
  (2012), pp.~1094--1121.

\bibitem{Fig:2007}
{\sc M.~A.~T. {Figueiredo}, R.~D. {Nowak}, and S.~J. {Wright}}, {\em Gradient
  projection for sparse reconstruction: Application to compressed sensing and
  other inverse problems}, IEEE Journal of Selected Topics in Signal
  Processing, 1 (2007), pp.~586--597.

\bibitem{Flinth:2019}
{\sc A.~Flinth and A.~Hashemi}, {\em Thermal source localization through
  infinite-dimensional compressed sensing}, ArXiv e-prints,  (2017).

\bibitem{Gholami:2016b}
{\sc A.~Gholami, J.~Hill, D.~Malhotra, and G.~Biros}, {\em {AccFFT}: {A}
  library for distributed-memory {FFT} on {CPU} and {GPU} architectures}, arXiv
  e-prints,  (2016).

\bibitem{Gholami:2016a}
{\sc A.~Gholami, A.~Mang, and G.~Biros}, {\em An inverse problem formulation
  for parameter estimation of a reaction-diffusion model of low grade gliomas},
  Journal of Mathematical Biology, 72 (2016), pp.~409--433.

\bibitem{Gholami:2017a}
{\sc A.~Gholami, A.~Mang, K.~Scheufele, C.~Davatzikos, M.~Mehl, and G.~Biros},
  {\em A framework for scalable biophysics-based image analysis}, in Proc
  ACM/IEEE Conference on Supercomputing, no.~19, 2017, pp.~19:1--19:13.

\bibitem{brats18}
{\sc A.~Gholami, S.~Subramanian, V.~Shenoy, N.~Himthani, X.~Yue, S.~Zhao,
  P.~Jin, G.~Biros, and K.~Keutzer}, {\em A novel domain adaptation framework
  formedical image segmentation}, Lecture Notes in Computer Science,  (2018).

\bibitem{Gong:2013}
{\sc P.~Gong, C.~Zhang, Z.~Lu, J.~Z. Huang, and J.~Ye}, {\em A general
  iterative shrinkage and thresholding algorithm for non-convex regularized
  optimization problems}, in Proceedings of the 30th International Conference
  on International Conference on Machine Learning - Volume 28, ICML'13,
  JMLR.org, 2013, pp.~II--37--II--45.

\bibitem{gooya2011deformable}
{\sc A.~Gooya, G.~Biros, and C.~Davatzikos}, {\em Deformable registration of
  glioma images using em algorithm and diffusion reaction modeling}, IEEE
  transactions on medical imaging, 30 (2011), pp.~375--390.

\bibitem{gooya-biros-davatzikos-e12}
{\sc A.~Gooya, K.~Pohl, M.~Bilello, L.~Cirillo, G.~Biros, E.~Melhem, and
  C.~Davatzikos}, {\em {GLISTR}: Glioma image segmentation and registration},
  IEEE Transactions on Medical Imaging, PP (2012), p.~1.

\bibitem{HawkinsDaarud:2013b}
{\sc A.~Hawkins-Daarud, S.~prudhomme, K.~G. van~der Zee, and J.~T. Oden}, {\em
  Bayesian calibration, validation, and uncertainty quantification of diffuse
  interface models of tumor growth}, J Math Biol, 67 (2013), pp.~1457--1485.

\bibitem{hogea2007modeling}
{\sc C.~Hogea, C.~Davatzikos, and G.~Biros}, {\em Modeling glioma growth and
  mass effect in {3D MR} images of the brain}, in Medical Image Computing and
  Computer-Assisted Intervention--MICCAI 2007, Springer, 2007, pp.~642--650.

\bibitem{Hogea:2007b}
{\sc C.~Hogea, C.~Davatzikos, and G.~Biros}, {\em Modeling glioma growth and
  mass effect in {3D} {MR} images of the brain}, in Lect Notes Comput Sc,
  vol.~4791, 2007, pp.~642--650.

\bibitem{hogea2008brain}
{\sc C.~Hogea, C.~Davatzikos, and G.~Biros}, {\em Brain-tumor interaction
  biophysical models for medical image registration}, SIAM Journal on
  Scientific Computing, 30 (2008), pp.~3050--3072.

\bibitem{Hogea:2008b}
{\sc C.~Hogea, C.~Davatzikos, and G.~Biros}, {\em An image-driven parameter
  estimation problem for a reaction-diffusion glioma growth model with mass
  effect}, J Math Biol, 56 (2008), pp.~793--825.

\bibitem{hormuth2018}
{\sc D.~A. Hormuth, S.~L. Eldridge, J.~A. Weis, M.~I. Miga, and T.~E.
  Yankeelov}, {\em Mechanically coupled reaction-diffusion model to predict
  glioma growth: Methodological details},  (2018), pp.~225--241.

\bibitem{Jaroudi2019}
{\sc R.~Jaroudi, G.~Baravdish, B.~T. Johansson, and F.~Astr\:{o}m}, {\em
  Numerical reconstruction of brain tumours}, Inverse Problems in Science and
  Engineering, 27 (2019), pp.~278--298.

\bibitem{jbabdi2005simulation}
{\sc S.~Jbabdi, E.~Mandonnet, H.~Duffau, L.~Capelle, K.~R. Swanson,
  M.~P{\'e}l{\'e}grini-Issac, R.~Guillevin, and H.~Benali}, {\em Simulation of
  anisotropic growth of low-grade gliomas using diffusion tensor imaging},
  Magnetic Resonance in Medicine, 54 (2005), pp.~616--624.

\bibitem{Khor:2019}
{\sc S.~Khoramian}, {\em An iterative thresholding algorithm for linear inverse
  problems with multi-constraints and its applications}, Applied and
  Computational Harmonic Analysis, 32 (2012), pp.~109 -- 130.

\bibitem{Kim:2007}
{\sc S.~{Kim}, K.~{Koh}, M.~{Lustig}, S.~{Boyd}, and D.~{Gorinevsky}}, {\em An
  interior-point method for large-scale$\ell_1$-regularized least squares},
  IEEE Journal of Selected Topics in Signal Processing, 1 (2007), pp.~606--617.

\bibitem{Knopoff:2013a}
{\sc D.~A. Knopoff, D.~R. Fern\'andez, G.~A. Torres, and C.~V. Turner}, {\em
  Adjoint method for a tumor growth {PDE}-constrained optimization problem},
  Computers \& Mathematics with Applications, 66 (2013), pp.~1104--1119.

\bibitem{Konukoglu:2010a}
{\sc E.~Konukoglu, O.~Clatz, P.~Y. Bondiau, H.~Delingette, and N.~Ayache}, {\em
  Extrapolating glioma invasion margin in brain magnetic resonance images:
  {S}uggesting new irradiation margins}, Medical Image Analysis, 14 (2010),
  pp.~111--125.

\bibitem{konukoglu2010image}
{\sc E.~Konukoglu, O.~Clatz, B.~H. Menze, B.~Stieltjes, M.-A. Weber,
  E.~Mandonnet, H.~Delingette, and N.~Ayache}, {\em Image guided
  personalization of reaction-diffusion type tumor growth models using modified
  anisotropic {Eikonal} equations}, Medical Imaging, IEEE Transactions on, 29
  (2010), pp.~77--95.

\bibitem{Le:2017a}
{\sc M.~L\^e, H.~Delingette, J.~Kalapathy-Cramer, E.~R. Gerstner, T.~Batchelor,
  J.~Unkelbach, and N.~Ayache}, {\em Personalized radiotherapy planning based
  on a computational tumor growth model}, Medical Imaging, IEEE Transactions
  on, 36 (2017), pp.~815--825.

\bibitem{Le:2015a}
{\sc M.~L\^e, H.~Delingette, J.~Kalpathy-Cramer, E.~R. Gerstner, T.~Batchelor,
  J.~Unkelbach, and N.~Ayache}, {\em Bayesian personalization of brain tumor
  growth model}, in Proc Medical Image Computing and Computer-Assisted
  Intervention, 2015, pp.~424--432.

\bibitem{Li:2014}
{\sc Y.~Li, S.~Osher, and R.~Tsai}, {\em Heat source identification based on l1
  constrained minimization}, Inverse Problems and Imaging, 8 (2014).

\bibitem{Mang:2014a}
{\sc A.~Mang}, {\em Methoden zur numerischen Simulation der Progression von
  Gliomen: Modellentwicklung, Numerik und Parameteridentifikation [in german;
  Methods for the numerical simulation of the progression of glioma: {M}odel
  design, numerics and parameter identification]}, Springer Fachmedien
  Wiesbaden, 2014.

\bibitem{Mang:2012a}
{\sc A.~Mang, T.~A. Schuetz, S.~Becker, A.~Toma, and T.~M. Buzug}, {\em Cyclic
  numerical time integration in variational non-rigid image registration based
  on quadratic regularisation}, in Proc Vision, Modeling and Visualization
  Workshop, 2012, pp.~143--150.

\bibitem{mang2012biophysical}
{\sc A.~Mang, A.~Toma, T.~A. Schuetz, S.~Becker, T.~Eckey, C.~Mohr,
  D.~Petersen, and T.~M. Buzug}, {\em Biophysical modeling of brain tumor
  progression: from unconditionally stable explicit time integration to an
  inverse problem with parabolic {PDE} constraints for model calibration},
  Medical Physics, 39 (2012), pp.~4444--4459.

\bibitem{Menze:2015a}
{\sc B.~H. Menze, A.~Jakab, S.~Bauer, J.~Kalpathy-Cramer, K.~Farahani, et~al.},
  {\em The multimodal brain tumor image segmentation benchmark ({BRATS})}, IEEE
  Transactions on Medical Imaging, 34 (2015), pp.~1993--2024.

\bibitem{menze2015multimodal}
{\sc B.~H. Menze, A.~Jakab, S.~Bauer, J.~Kalpathy-Cramer, K.~Farahani,
  J.~Kirby, Y.~Burren, N.~Porz, J.~Slotboom, R.~Wiest, et~al.}, {\em The
  multimodal brain tumor image segmentation benchmark (brats)}, IEEE
  transactions on medical imaging, 34 (2015), pp.~1993--2024.

\bibitem{Mi:2014a}
{\sc H.~Mi, C.~Petitjean, B.~Dubray, P.~Vera, and S.~Ruan}, {\em Prediction of
  lung tumor evolution during radiotherapy in individual patients with {PET}},
  Medical Imaging, IEEE Transactions on, 33 (2014), pp.~995--1003.

\bibitem{More:1994a}
{\sc J.~J. Mor{\'e} and D.~J. Thuente}, {\em Line search algorithms with
  guaranteed sufficient decrease}, ACM Trans. Math. Softw., 20 (1994),
  pp.~286--307.

\bibitem{murray1989mathematical}
{\sc J.~D. Murray}, {\em Mathematical biology}, Springer-Verlag, New York,
  1989.

\bibitem{NEEDELL2009301}
{\sc D.~Needell and J.~Tropp}, {\em Cosamp: Iterative signal recovery from
  incomplete and inaccurate samples}, Applied and Computational Harmonic
  Analysis, 26 (2009), pp.~301 -- 321.

\bibitem{Needell:2010}
{\sc D.~{Needell} and R.~{Vershynin}}, {\em Signal recovery from incomplete and
  inaccurate measurements via regularized orthogonal matching pursuit}, IEEE
  Journal of Selected Topics in Signal Processing, 4 (2010), pp.~310--316.

\bibitem{Nocedal:2006a}
{\sc J.~Nocedal and S.~J. Wright}, {\em Numerical Optimization}, Springer, New
  York, New York, US, 2006.

\bibitem{pham2012density}
{\sc K.~Pham, A.~Chauviere, H.~Hatzikirou, X.~Li, H.~M. Byrne, V.~Cristini, and
  J.~Lowengrub}, {\em Density-dependent quiescence in glioma invasion:
  instability in a simple reaction--diffusion model for the
  migration/proliferation dichotomy}, Journal of biological dynamics, 6 (2012),
  pp.~54--71.

\bibitem{Quiroga:2015a}
{\sc A.~A.~I. Quiroga, D.~Fern\'andez, G.~A. Torres, and C.~V. Turner}, {\em
  Adjoint method for a tumor invasion {PDE}-constrained optimization problem in
  {2D} using adaptive finite element method}, Applied Mathematics and
  Computation, 270 (2015), pp.~358--368.

\bibitem{akbari-e18}
{\sc S.~Rathore, H.~Akbari, J.~Doshi, C.~A. Davatzikos, et~al.}, {\em Radiomic
  signature of infiltration in peritumoral edema predicts subsequent recurrence
  in glioblastoma: implications for personalized radiotherapy planning},
  Journal of Medical Imaging, 5 (2018), pp.~5 -- 5 -- 10.

\bibitem{rekik2013tumor}
{\sc I.~Rekik, S.~Allassonni{\`e}re, O.~Clatz, E.~Geremia, E.~Stretton,
  H.~Delingette, and N.~Ayache}, {\em Tumor growth parameters estimation and
  source localization from a unique time point: Application to low-grade
  gliomas}, Computer Vision and Image Understanding, 117 (2013), pp.~238--249.

\bibitem{REKIK2013238}
{\sc I.~Rekik, S.~Allassonnière, O.~Clatz, E.~Geremia, E.~Stretton,
  H.~Delingette, and N.~Ayache}, {\em Tumor growth parameters estimation and
  source localization from a unique time point: Application to low-grade
  gliomas}, Computer Vision and Image Understanding, 117 (2013), pp.~238 --
  249.

\bibitem{rockne-e19}
{\sc R.~C. Rockne, A.~Hawkins-Daarud, K.~R. Swanson, J.~P. Sluka, J.~A.
  Glazier, P.~Macklin, D.~Hormuth, A.~M. Jarrett, E.~A.~B. da~Fonseca~Lima,
  J.~Oden, G.~Biros, T.~E. Yankeelov, K.~Curtius, I.~A. Bakir, D.~Wodarz,
  N.~Komarova, L.~Aparicio, M.~Bordyuh, R.~Rabadan, S.~Finley, H.~Enderling,
  J.~J. Caudell, E.~G. Moros, A.~R.~A. Anderson, R.~Gatenby, A.~Kaznatcheev,
  P.~Jeavons, N.~Krishnan, J.~Pelesko, R.~R. Wadhwa, N.~Yoon, D.~Nichol,
  A.~Marusyk, M.~Hinczewski, and J.~G. Scott}, {\em The 2019 mathematical
  oncology roadmap}, Physical Biology,  (2019).

\bibitem{saut2014multilayer}
{\sc O.~Saut, J.-B. Lagaert, T.~Colin, and H.~M. Fathallah-Shaykh}, {\em A
  multilayer grow-or-go model for {GBM}: effects of invasive cells and
  anti-angiogenesis on growth}, Bulletin of mathematical biology, 76 (2014),
  pp.~2306--2333.

\bibitem{Scheufele:2019b}
{\sc K.~Scheufele}, {\em {Coupling Schemes and Inexact Newton for Multi-Physics
  and Coupled Optimization Problems}}, doctoral thesis, University of
  Stuttgart, Faculty of Computer Science, Electrical Engineering, and
  Information Technology, Germany, February 2019.

\bibitem{Scheufele:2017a}
{\sc K.~Scheufele, A.~Mang, A.~Gholami, C.~Davatzikos, G.~Biros, and M.~Mehl},
  {\em Coupling brain-tumor biophysical models and diffeomorphic image
  registration}, Computer Methods in Applied Mechanics and Engineering, 347
  (2019), pp.~533--567.

\bibitem{subramanian-gholami-biros19}
{\sc S.~Subramanian, A.~Gholami, and G.~Biros}, {\em Simulation of glioblastoma
  growth using a {3D} multispecies tumor model with mass effect}, Journal of
  Mathematical Biology,  (2019).

\bibitem{swanson2000quantitative}
{\sc K.~Swanson, E.~Alvord, and J.~Murray}, {\em A quantitative model for
  differential motility of gliomas in grey and white matter}, Cell
  Proliferation, 33 (2000), pp.~317--330.

\bibitem{swanson2002virtual}
{\sc K.~R. Swanson, E.~Alvord, and J.~Murray}, {\em Virtual brain tumours
  (gliomas) enhance the reality of medical imaging and highlight inadequacies
  of current therapy}, British Journal of Cancer, 86 (2002), pp.~14--18.

\bibitem{Swanson:2011a}
{\sc K.~R. Swanson, R.~C. Rockne, J.~Claridge, M.~A. Chaplain, E.~C. Alvord,
  and A.~R. Anderson}, {\em Quantifying the role of angiogenesis in malignant
  progression of gliomas: In silico modeling integrates imaging and histology},
  Cancer Res, 71 (2011), pp.~7366--7375.

\bibitem{Swanson2008}
{\sc K.~R. Swanson, R.~C. Rostomily, and J.~Alvord, E~C}, {\em A mathematical
  modelling tool for predicting survival of individual patients following
  resection of glioblastoma: a proof of principle}, British journal of cancer,
  98 (2008), pp.~113--119.

\bibitem{Tropp:2007}
{\sc J.~A. {Tropp} and A.~C. {Gilbert}}, {\em Signal recovery from random
  measurements via orthogonal matching pursuit}, IEEE Transactions on
  Information Theory, 53 (2007), pp.~4655--4666.

\bibitem{Tropp10}
{\sc J.~A. {Tropp} and S.~J. {Wright}}, {\em Computational methods for sparse
  solution of linear inverse problems}, Proceedings of the IEEE, 98 (2010),
  pp.~948--958.

\bibitem{Wong:2017a}
{\sc K.~C.~L. Wong, R.~M. Summers, E.~Kebebew, and J.~Yoa}, {\em Pancreatic
  tumor growth prediction with elastic-growth decomposition, image-derived
  motion, and {FDM-FEM} coupling}, Medical Imaging, IEEE Transactions on, 36
  (2017), pp.~111--123.

\end{thebibliography}
\bibliographystyle{siam}
\end{document}